\documentclass[fleqn,usenatbib,onecolumn]{mnras}

\usepackage[T1]{fontenc}

\DeclareRobustCommand{\VAN}[3]{#2}
\let\VANthebibliography\thebibliography
\def\thebibliography{\DeclareRobustCommand{\VAN}[3]{##3}\VANthebibliography}

\usepackage{graphicx}	
\usepackage{amsmath}	
\usepackage{amssymb}	
\usepackage{bm} 
\usepackage{mathtools} 
\usepackage{circuitikz} 
\usetikzlibrary{math, decorations.pathmorphing, patterns}

\newcommand{\mathsout}[1]
{\bgroup\mathchoice
	{\sbox0{$\displaystyle{#1}$}%
		\usebox0\hspace{-\wd0}%
		\rule[0.5\ht0-0.5\dp0-.5pt]{\wd0}{0.6pt}}%
	{\sbox0{$\textstyle{#1}$}%
		\usebox0\hspace{-\wd0}%
		\rule[0.5\ht0-0.5\dp0-.5pt]{\wd0}{0.6pt}}%
	{\sbox0{$\scriptstyle{#1}$}%
		\usebox0\hspace{-\wd0}%
		\rule[0.5\ht0-0.5\dp0-.5pt]{\wd0}{0.6pt}}%
	{\sbox0{$\scriptscriptstyle{#1}$}%
		\usebox0\hspace{-\wd0}%
		\rule[0.5\ht0-0.5\dp0-.5pt]{\wd0}{0.6pt}}%
	\egroup}

\usepackage{newtxtext,newtxmath}

\title[Gravitational radiation back-reaction from f-modes]{Gravitational radiation back-reaction from  f-modes on neutron stars}

\author[G. Yim \& D. I. Jones]{
Garvin Yim\thanks{E-mail: g.yim@soton.ac.uk} and
D. I. Jones\thanks{E-mail: d.i.jones@soton.ac.uk} 
\\
Mathematical Sciences and STAG Research Centre, University of Southampton, Southampton SO17 1BJ, UK
}

\date{Accepted XXX. Received YYY; in original form ZZZ}

\pubyear{2021}

\begin{document}
\label{firstpage}
\pagerange{\pageref{firstpage}--\pageref{lastpage}}
\maketitle

\begin{abstract}
The problem of the gravitational radiation damping of neutron star fundamental ($f$) mode oscillations has received considerable attention. Many studies have looked at the stability of such oscillations in rapidly rotating stars, calculating the growth/decay rate of the mode amplitude. In this paper, we look at the relatively neglected problem of the radiation reaction on the spin of the star. We specialise greatly to the so-called Kelvin modes: the modes of oscillation of (initially) non-rotating incompressible stars. We find the unexpected result that the excitation of a mode of angular momentum $\delta J$ on an initially non-rotating star ends up radiating an angular momentum $2\delta J$ to infinity, leaving the star itself with a bulk angular momentum of $-\delta J$. This result is interesting in itself, and also will have implications for the angular momentum budgets of spinning down neutron stars, should such modes be excited.

\end{abstract}




\section{Introduction}
\label{section: introduction}

The oscillation modes of compact objects, whether it be a black hole or neutron star (NS), are of great interest to the rapidly developing field of gravitational wave (GW) astronomy \citep{kokkotasSchmidt1999, andersson2021}. With the discovery of GWs from inspiralling compact binaries \citep{abbottetal2019GWTC1, abbottetal2021GWTC2}, there is now an enhanced vigour to detect other forms of GWs, namely, stochastic, continuous and bursts. Any such detection would be the first of its kind.

The hope is that, someday in the future, detections of GWs from oscillating compact objects would shed light on the complex physics that govern these objects. For NSs, this includes information about their concealed interiors \citep[e.g.][]{anderssonKokkotas1998, kokkotasApostolatosAndersson2001, krugerKokkotas2020} and moreover, these GWs could provide additional tests of gravity \citep[e.g.][]{sotaniKokkotas2004}.

There exists a multitude of different oscillation modes, including the $p$, $g$ and $r$-modes to name a few, but here, we focus only on the fundamental ($f$) modes. $f$-modes can be thought of as ``bulk'' perturbations of a NS, with no radial nodes. Essentially, $f$-modes correspond to changes to the entire shape, according to some displacement vector, $\xi$. Our analysis concerns itself entirely with the \textit{gravitational radiation reaction} problem for these modes, i.e.~the way in which GW energy loss damps the oscillation, and exerts a torque on the NS.

We provide a specific analysis for the simplest case of an initially non-rotating, uniformly-dense and incompressible NS, in Newtonian gravity. As a result, the only modes that can be excited are the $f$-modes and when combined with the above assumptions, are called the \textit{Kelvin modes}, named after Lord Kelvin who first discovered them \citep{thomson1863}. The non-rotating assumption is justified for slowly rotating NSs that rotate much slower than their break-up frequency. Any corrections would be of the order $\mathcal{O}(\Omega/\Omega_*)$, where $\Omega_*$ is the break-up frequency. Moreover, we focus on the $l=2$ modes as these are the lowest order multipoles that emit GWs and contribute more to GW emission than higher order multipoles \citep{thorne1980}. These assumptions allow our work to be completely analytic. In time, these should be relaxed to ensure our findings here remain valid.

The effect of the gravitational radiation reaction on more realistic stellar configurations has in fact been considered in detail, via numerical calculations, including the work of \citet{lindblom1986}, \citet{ipserLindblom1991}, \citet{donevaetal2013} and \citet{krugerKokkotas2020}. However, these analyses have concentrated on the stability of modes in rapidly rotating stars with regard to the GW-driven Chandrasekhar-Friedman-Schutz (CFS) instability \citep{chandrasekhar1970, friedmanschutz1978, friedmanschutz1978b}. Our goal is different -- it is to understand the effect of radiation reaction on non-rotating (and therefore CFS-stable) stars, looking not only at the mode amplitude, but also at the torque exerted on the star, and its implication for the spin. In fact, the damping of modes due to radiation reaction have been calculated previously by \cite{chau1967} for axisymmetric modes and \cite{detweiler1975}, based on \cite{thorne1969iv}, for non-axisymmetric modes, but these calculations only consider energy conservation and not the full problem which also includes angular momentum. As such, our calculation can be thought of as the non-rotating Kelvin-mode equivalent of the \citet{owenetal1998} calculation of $r$-mode evolution. As far as we are aware, this rather simple sort of calculation has not been reported before. 


Our calculations are largely based on the work of \citet{friedmanschutz1978, friedmanschutz1978b} who presented a formalism for second order perturbation theory which is necessary for calculating mode energies and angular momenta. By conserving these quantities and ensuring their decay time-scales match, we show that there must be a GW back-reaction on an oscillating NS, where the emission of GWs from a prograde (retrograde) mode causes the NS to rotate in the retrograde (prograde) direction. This is somewhat counter-intuitive as one would think GW emission only acts to damp the mode and have no effect of the spin, though we show this is not true. Interestingly, the aforementioned GW back-reaction can also be seen from a simple toy model involving masses and springs. Further details of this can be found in Appendix~\ref{appendix: toy model}.

The arrangement of this paper is as follows. In Section~\ref{section: mode eigenfunctions and frequencies}, we introduce the properties of the Kelvin modes. In Section~\ref{section: mode energies and angular momenta}, we calculate the physical energy and angular momentum of the modes.  In Section~\ref{section: gravitational wave emission}, we find the rate of change of these quantities due to GW emission. We use this to show the effect of radiation reaction cannot simply be to damp the mode -- surprisingly (to us), the star must gain a rotational angular momentum.
In Section~\ref{section: canonical mode energies and angular momenta}, we use the formalism of \citet{friedmanschutz1978, friedmanschutz1978b} to compute the canonical energies and angular momenta of the modes, as well as their time derivatives. This allows us to compute the full response of the star to radiation reaction in Section~\ref{section: gravitational wave back-reaction}.  Finally, in Section~\ref{section: conclusions}, we finish with some comments as well as our conclusions.

We make extensive use of equations from \citet{friedmanschutz1978} and \citet{friedmanschutz1978b}, which we refer to using the shorthand (FSa...) and (FSb...) respectively.

\section{Properties of Kelvin modes}
\label{section: mode eigenfunctions and frequencies}

We begin by introducing the displacement vector/eigenfunction of the Kelvin modes, which are $f$-modes under the assumptions of no rotation, uniform density and incompressibility. The (real) displacement vector, $\bm{\xi}$, relates the perturbed position of some fluid element, $\tilde{\textbf{r}}$, to its unperturbed position,~$\textbf{r}$, via
\begin{align}
\label{unperturbed_to_perturbed}
\tilde{\textbf{r}}(\textbf{r},t) \equiv \textbf{r} + \bm{\xi}(\textbf{r},t)~.
\end{align}
In spherical coordinates, the Kelvin mode eigenfunction is (see Ch.~13.1 of \citet{andersson2019book} for a derivation)
\begin{align}
\label{real_xi_general_l_Re}
{\bm{\xi}} (r, \theta, \phi, t) &= \Re \left\{ \frac{\alpha_{lm}}{R^{l-2}} \bm{\nabla} [r^l Y_{lm}(\theta, \phi)] e^{i\omega_{l}t} \right\}~,
\end{align}
where $\Re$ means to take the real part, $\alpha_{lm}$ is our small dimensionless amplitude parameter ($\alpha_{lm} \ll 1$), $R$ is the unperturbed NS radius, $\bm{\nabla}$ is the gradient operator, $Y_{lm}$ are the spherical harmonics with degree $l$ and order $m$, and $\omega_{l}$ is the mode frequency, which, when squared, is
\begin{align}
\label{kelvin_mode_frequency_general_l}
\omega_{l}^2 = \frac{8 \pi G \bar{\rho}}{3} \frac{l(l-1)}{2l+1}~,
\end{align}
where $\bar{\rho} = \frac{3 M}{4 \pi R^3}$ is the density of our uniform NS. The scale factor of $R^{l-2}$ in the denominator of $\xi$ means $\alpha_{lm}$ is dimensionless for all $l$. Note that these modes, by construction, satisfy the incompressibility condition
\begin{align}
\label{incompressibility}
\bm{\nabla} \cdot {\bm{\xi}} = 0~.
\end{align}
In explicitly real form, we have
\begin{align}
\label{real_xi_general_l}
\bm{\xi} (r, \theta, \phi, t) = \alpha_{lm} \frac{r^{l-1}}{R^{l-2}}  \left[ l Y_{lm}(\theta, 0) \cos(m\phi + \omega_{l} t) \bm{e}_r 
+  
\frac{d Y_{lm}(\theta, 0)}{d\theta} \cos(m\phi + \omega_{l} t) \bm{e}_\theta 
- 
\frac{m }{\sin\theta} Y_{lm}(\theta, 0) \sin(m\phi + \omega_{l} t) \bm{e}_\phi\right]~.
\end{align}

We have taken the trouble to write the operator $\Re$ explicitly, for two reasons. Firstly, we will be concerned with calculating various energies and angular momenta which are second order in $\xi$, computation of which requires one to first take the real parts. Also, some of the results from \citet{friedmanschutz1978,friedmanschutz1978b} that we will use later explicitly require \emph{complex} $\xi$, which we will denote as $\tilde\xi$. As we will see later, the complex $\tilde\xi$ require a slightly different normalisation from the real $\xi$. Hence our efforts to distinguish carefully between the two. 


Non-axisymmetric modes ($m\ne0$) propagate around the NS at pattern speed
\begin{align}
\label{pattern_speed}
\omega_\text{p} \equiv - \frac{\omega_{l}}{m}~,
\end{align}
which comes from tracking a fixed phase of the mode, i.e.~from $\frac{d}{dt}\Phi = \frac{d}{dt}(m\phi + \omega_{l} t) = 0$. This means that modes with negative $m$ propagate in the prograde (positive) direction and vice versa for positive $m$.

We will also need the (Eulerian) changes to the pressure, $\delta P$, and (internal) gravitational potential, $\delta \Phi$, of the Kelvin modes a little later. The derivation of $\delta P$ and $\delta \Phi$ is an essential part of, and naturally comes from, the derivation of the Kelvin modes \citep[e.g. Ch.~13.1 of][]{andersson2019book}. We will not repeat the calculation here. Instead, we take the results and tailor them for our small parameter, $\alpha_{lm}$, which gives
\begin{align}
\label{delta_P}
\delta P &= \frac{4 \pi G \bar{\rho}^2 l}{3} \alpha_{lm} \frac{r^l}{R^{l-2}} Y_{lm}(\theta, 0) \cos(m \phi + \omega_{l} t)~, \\
\label{delta_Phi}
\delta \Phi &= - \frac{4 \pi G \bar{\rho} l}{2l+1} \alpha_{lm} \frac{r^l}{R^{l-2}} Y_{lm}(\theta, 0) \cos(m \phi + \omega_{l} t)~.
\end{align}
As it will also be needed later, we will also write down the perturbation in the gravitational potential outside of the star (i.e.~for $r > R$)
\begin{align}
\label{delta_Phi_ext}
\delta \Phi_{\rm ext} &= - \frac{4 \pi G \bar{\rho} l}{2l+1} \alpha_{lm} \frac{R^{l+3}}{r^{l+1}} Y_{lm}(\theta, 0) \cos(m \phi + \omega_{l} t)~.
\end{align}

\section{Mode energies and angular momenta}
\label{section: mode energies and angular momenta}

With a view to calculating mode damping time-scales, we now proceed to calculate mode energies and angular momenta for quadrupolar ($l=2$) modes, accurate to second order in $\alpha_{2, m}$.  These are the ``physical'' mode energies and angular momenta, that follow from the elementary definitions of energy and angular momenta, with the label ``physical'' being used to distinguish them from the ``canonical'' energies and angular momenta that will be considered later (Section \ref{section: canonical mode energies and angular momenta}), consistent with the nomenclature of  \citet{friedmanschutz1978}.

\subsection{Mode energies}
\label{subsection: mode energies}

In their Appendix B, \citet{friedmanschutz1978} give all the formulae that are needed to compute mode energies accurate to second order, using only the first order Lagrangian perturbations, $\xi$. We simply need to evaluate the relevant formulae for Kelvin modes. We give a brief sketch of these calculations and present the results for $l=2$ modes. 

The total energy perturbation is given as the sum of kinetic, internal, and gravitational pieces (FSaB59)
\begin{align}
\delta E = \delta T + \delta U + \delta W~.
\end{align}
We give a summary in Table~\ref{table:energies} of each of these pieces, for the cases $l=2$, $m=0$ and $l=2$, $m \neq 0$.

\begin{table}
 \begin{minipage}{165mm}
 \caption{Table of contributions to the total mode energy $\delta E$, in units of $\beta \equiv \alpha_{2,m}^2 \bar\rho \omega_2^2 R^5$. Starting with column 2, we have perturbation in kinetic energy $\delta T$, internal energy $\delta U$, gravitational energy $\delta W$, ``total potential energy'' $\delta V \equiv \delta U + \delta W$, and total energy $\delta E = \delta T + \delta V$.}
 \label{table:energies}
\begin{tabular}{@{}llllll}
\hline
& $\delta T / \beta$ & $\delta U / \beta$ & $\delta W / \beta$ & $\delta V / \beta$ & $\delta E / \beta$ \\
\hline
$l=2$, $m=0$ & $\sin^2(\omega_2 t)$ & $\frac{5}{4} \cos^2(\omega_2 t)$ & $-\frac{1}{4} \cos^2(\omega_2 t)$ &
$\cos^2(\omega_2 t)$ & $1$ \\ [-6pt]
\\
$l=2$, $m = \pm 1, \,\pm 2$ & $\frac{1}{2}$ & $\frac{5}{8}$ & $-\frac{1}{8}$ & $\frac{1}{2}$ & $1$ \\
\hline
\end{tabular}
\end{minipage}
\end{table}

The kinetic energy perturbation is given by (FSaB43), which for a non-rotating star takes the extremely simple form
\begin{align}
\delta T = \frac{1}{2} \int_V \rho \dot\xi^i \dot\xi_i \, dV~.
\end{align}
Substituting for $\xi$ using equation~(\ref{real_xi_general_l}) we obtain the results given in the second column of Table \ref{table:energies}. Note that for $m=0$, we have $\delta T \propto \sin^2(\omega_2 t)$, while for $m \neq 0$ we have $\delta T = \text{constant}$. Both of these results were to be expected. For $m=0$, the oscillation changes the shape of the star, between an axisymmetric prolate shape to an axisymmetric oblate one and back again, once per mode period; the kinetic energy will be zero at each of these two extremes, and positive inbetween. For $m \neq 0$, the mode instead results in a perturbed shape that rotates at the pattern speed, so that the stellar configurations at different times is related by a simple rotation about the $z$-axis; it follows that $\delta T$ (and also $\delta U$ and $\delta W$) must be constant in time.

The internal energy piece is given by the $\nabla_i \xi^i = 0$ form of (FSaB48)
\begin{align}
\delta U = \frac{1}{2} \int_V \xi^i \xi^j \nabla_i \nabla_j P~dV~,
\end{align}
where $P(r)$ is the background (i.e.\ unperturbed) pressure. There is a small subtlety involved in performing this integral. The quantity $\nabla_j P$ is the net pressure force per unit volume in the background star. The non-vanishing of the mass per unit volume at the surface results in step-like behaviour in $\nabla_j P$ at the stellar surface, that can be described via a Heaviside step function
\begin{align}
\nabla_j P(r) = -\frac{4}{3} \pi G \bar\rho^2 r_j H(R-r)~,
\end{align}
where $r_j$ is the position vector. When differentiated, the step function results in the appearance of a delta function
\begin{align}
\nabla _i \nabla_j P(r) = -\frac{4}{3} \pi G \bar\rho^2 [\delta_{ij} H(R-r) - r_j \hat r_i \delta(r-R)]~,
\end{align}
where $\hat r_i$ is the radial unit vector. The integral to be computed therefore contains both a volume and surface term
\begin{align}
\delta U = -\frac{2}{3} \pi G \bar\rho^2 \int_V \xi^i \xi_i \, dV + \frac{2}{3} \pi G \bar\rho^2 R \int_{\partial V} [\xi^r(R)]^2 \, dS~.
\end{align}
The results of this computation are given in the third column of Table~\ref{table:energies}, showing positive non-zero values for $\delta U$.

It may seem surprising that the perturbation of the internal energy is not exactly zero for the modes we consider, which are, after all, supposed to correspond to perturbation of an incompressible background star. We attribute the existence of this non-zero perturbation in internal energy to the fact that the mode solutions of Section~\ref{section: mode eigenfunctions and frequencies} were obtained by solving the equations of motion to \emph{first} order in $\xi$. As such, they satisfy the incompressibility condition of equation~(\ref{incompressibility}), i.e.~$\nabla_i \xi^i = 0$, but only to first order. To second order, there will be a non-zero Lagrangian perturbation in the density, as is made clear by (FSaB32)
\begin{align}
\frac{\Delta \rho}{\bar\rho} = -\nabla_i \xi^i 
+ \frac{1}{2} (\nabla_i\xi^i \nabla_j\xi^j + \nabla_i\xi^j \nabla_j\xi^i) + {\mathcal O}(\xi^3)~.
\end{align}
The last of the three terms on the right hand side is non-zero for the mode solutions of equation~(\ref{real_xi_general_l}), giving the net compression that leads to the positive $\delta U$ values we have computed.

The perturbation in gravitational energy is given by equation~(FSaB56)
\begin{align}
\delta W = \int_V \left[ \rho \xi^i\nabla_i \Phi + \rho\xi^i\nabla_i\delta\Phi + \frac{1}{8\pi G} \nabla_i\delta\Phi \nabla^i\delta\Phi
+ \frac{1}{2} \rho \xi^i \xi^j \nabla_i \nabla_j\Phi \right] ~ dV~.
\end{align}
Friedman \& Schutz obtain this result by integrating by parts several times, exploiting the fact that surface terms, evaluated at infinity, are zero (see (FSaB49)~--~(FSaB53)). It follows that the domain of integration $V$ itself extends to infinity, so one must include the contribution from \emph{outside} the star when evaluating the third term on the right hand side, making use of equation~(\ref{delta_Phi}) inside the star, and equation~(\ref{delta_Phi_ext}) outside. Performing the calculations, we obtain the results given in the fourth column of Table~\ref{table:energies}.

In the fifth column of Table~\ref{table:energies}, we give the sum of the internal and gravitational energy perturbations 
\begin{align}
\delta V \equiv \delta U + \delta W~.
\end{align}
This can be thought of as a potential energy, which when summed with the kinetic term gives the full energy perturbation, $\delta E$, as recorded in the final column of Table~\ref{table:energies}. For all $m$, the result can be written as
\begin{equation}
\label{eq:delta_E}
\delta E = \alpha_{2,m}^2 \bar\rho \omega_2^2 R^5~.
\end{equation}

We can perform several checks of these results. Firstly, for $m=0$, we can see that the $\delta T$ and $\delta V$ terms are of equal amplitude, representing an exchange between kinetic and potential energies, with constant total energy. For all $m$, we can verify that our results are consistent with the ``virial equation for perturbations'', given by equation~(6.6.14) of \citet{shapiroTeukolsky1983}
\begin{align}
\label{virial_equation}
\frac{1}{4} \frac{d^2}{dt^2} \delta I = \delta T - \delta V~,
\end{align}
where $\delta I$ is the trace of the mass quadrupole moment tensor caused by the perturbations. We postpone calculation of this tensor until Section~\ref{subsection: energy considerations}, where it is required to compute the GW emission, but note here that this virial equation is indeed satisfied by our solutions.

For $m=0$, we can also compare our result for $\delta E$ with what is written in \citet{chau1967}, in his equation~(17). Our result for $\delta E$ is a factor of $2$ larger than the result he uses. In fact, Chau does not compute $\delta E$. Instead, he takes a result from \citet{rayleigh1945}, but comparing with this original source, Chau seems to have incorrectly copied Rayleigh's result. Specifically, it appears Chau only considers the kinetic energy term, given by equation~(7) of \S 264 of \citet{rayleigh1945}, and replaces a $\cos^2$ term with a factor of $1/2$ (perhaps with the mindset of taking an average) when in fact, the inclusion of potential energies introduces a $\sin^2$ term which combines with the kinetic energy without the need to take an average.

\subsection{Mode angular momenta}
\label{subsection: mode angular momenta}

\citet{friedmanschutz1978b} also give all the formulae that are needed to compute mode angular momenta accurate to second order, using only the first order Lagrangian perturbations, $\xi$. 

Combining equations~(FSa61), (FSa47) and (FSa38) for $v^i=0$ we have
\begin{equation}
\delta J = - \int_V \rho \left[ \phi^i \Delta v_i + \frac{\partial \xi^i}{\partial \phi} \frac{\partial{\xi_i}}{\partial t} \right] \, dV~,
\end{equation} 
where $\phi^i = r \sin\theta (\bm{e}_\phi)^i$ and $\Delta v_i$ is the Lagrangian perturbation in velocity, accurate to second order in $\xi$, as given in equation~(FSaB40)
\begin{equation}
\Delta v_i = \dot\xi_i + \dot\xi^j \nabla_i \xi_j~.
\end{equation}
Using the $l=2$ form of $\xi$ from equation~(\ref{real_xi_general_l}), we find
\begin{equation}
\label{eq:delta_J}
\delta J = -\frac{1}{2} m \alpha_{2,m}^2 \bar\rho \omega_2 R^5~,
\end{equation}
valid for all $m$.

Recall our convention for the mode phase, $(m\phi + \omega_{l}t)$. Consistent with this, we see that modes which propagate in the positive mathematical sense ($m<0$) have $\delta J > 0$, while modes that propagate in the negative mathematical sense ($m>0$) have $\delta J < 0$, and $m=0$ ``prolate/oblate'' modes have $\delta J=0$.

\section{Gravitational wave emission}
\label{section: gravitational wave emission}

In Section~\ref{section: mode energies and angular momenta}, we calculated the energy and angular momenta carried by the $l=2$ Kelvin modes, neglecting dissipation. With a view to calculating the effect of gravitational radiation reaction on these modes, we will now compute the rate at which they radiate energy and angular momentum via GWs. The $m=0$ case was first studied by \citet{chau1967} and the $m \neq 0$ case by \citet{detweiler1975}. It is for $m \neq 0$ that we find our surprising result.



In Section~\ref{subsection: energy considerations}, we calculate the rate of change of energy due to GW emission. In Section~\ref{subsection: angular momentum considerations}, we calculate the rate of change of angular momentum. In Section~\ref{subsection: GW_timescales}, we discuss the significance of these results.


\subsection{Rate of change of energy}
\label{subsection: energy considerations}

For $l=2$, the rate of change of energy due to GW emission, $\dot{E}_\text{GW}$, is given by the standard quadrupole formula
\begin{align}
\label{GW_luminosity_quadrupole_formula}
\dot{E}_\text{GW} = \frac{1}{5} \frac{G}{c^5} \left\langle \mathsout{\dddot{I}}_{ij} \mathsout{\dddot{I}}^{~ij} \right\rangle~,
\end{align}
where the dots represent time derivatives in the inertial frame and the angled brackets represent an average over several wavelengths/periods \citep[e.g.][]{misnerThorneWheeler1973book, andersson2019book}. 
$\mathsout{I}_{ij}$ is the trace-reduced mass quadrupole moment tensor, defined by 
\begin{align}
\mathsout{I}_{ij} \equiv \int_{V} \rho \left( x_i x_j - \frac{1}{3} \delta_{ij} x^k x_k \right) dV~,
\end{align}
where $x^i$ is the position vector in a Cartesian basis. Note that $\dot{E}_\text{GW}$ is positive when GWs carry energy away from the system resulting in the system's energy decreasing. 

In practice, one would first calculate the mass quadrupole moment tensor
\begin{align}
\label{mass_quadrupole_moment_definition}
I_{ij} \equiv \int_{V} \rho x_i x_j dV~,
\end{align}
before reducing by its trace to give $\mathsout{I}_{ij}$, i.e.
\begin{align}
\label{reducing_by_trace}
\mathsout{I}_{ij} = I_{ij} - \frac{1}{3} \delta_{ij} I^k_{~k}~~,
\end{align}
such that $\text{Tr}(\mathsout{I}_{ij}) = 0$. It is useful to decompose the mass quadrupole moment tensor into a term related to the background, which is spherical and constant in time, and a time-dependent term which is caused by the perturbation
\begin{align}
{I}_{ij} = I_{\text{sph}} \delta_{ij} + \delta I_{ij} ~~ \rightarrow ~~ \dot{I}_{ij} = \delta \dot{I}_{ij} ~~ \rightarrow ~~ \mathsout{\dot{I}}_{ij} = \delta \mathsout{\dot{I}}_{ij}~,
\end{align}
and similarly for higher time derivatives. From this, we see that once we obtain $\delta I_{ij}$, we can reduce it by its trace, differentiate with respect to time an appropriate number of times and then substitute into equation~(\ref{GW_luminosity_quadrupole_formula}) to get $\dot{E}_\text{GW}$. So, our task now is to find $\delta I_{ij}$.

We can use the general rule for perturbing integral quantities where the mass density appears in the integrand
\begin{align}
\label{perturbing_integral_quantity}
\delta \int_{V} \rho Q dV = \int_{V} \rho \Delta Q dV~,
\end{align}
(FSaB12) for some fluid variable $Q$ (see also \textsection15 of \citet{chandrasekhar1969book} for a similar result but using the covariant derivative definition of the Lagrangian change, as opposed to the more natural definition based on Lie derivatives used by Friedman \& Schutz). We obtain
\begin{align}
\label{mass_quadrupole_moment_perturbations_definition}
\delta {I}_{ij} = \delta \int_{V} \rho x_i x_j dV = \int_{V} \rho \Delta (x_i x_j) dV = \bar{\rho} \int_{V} (\xi_i x_j + x_i \xi_j + \xi_i \xi_j) dV~,
\end{align}
which is exact and where $\xi$ is now in a Cartesian basis. Note that the Lagrangian change to the position vector is exactly $\xi$ by definition so there are no higher order terms to consider. We need only the first order terms of $\delta {I}_{ij}$ to calculate the rate of change in GW energy and angular momentum accurate to second order, which are quadratic in (the time derivatives of) $\delta {I}_{ij}$, but we need the second order term to test the virial equation stated earlier in equation~(\ref{virial_equation}).

Substituting $\xi$ from equation~(\ref{real_xi_general_l}) into equation~(\ref{mass_quadrupole_moment_perturbations_definition}) and after some algebra, we find 
\begin{align} 
\label{mass_quadrupole_2_0}
\delta I^{2,0}_{ij} &= \frac{1}{15} \alpha_{2,0} \bar{\rho} R^5 \cos(\omega_{2}t) \left(~
\begin{matrix}
-4\sqrt{5\pi} + 5 \alpha_{2,0}\cos(\omega_{2}t) & 0 & 0 \\ 
0 & -4\sqrt{5\pi} + 5 \alpha_{2,0}\cos(\omega_{2}t) & 0\\
0 & 0 & 4(2\sqrt{5\pi} + 5 \alpha_{2,0}\cos(\omega_{2}t)) \\
\end{matrix}
~\right)~, \\
\delta I^{2,\pm 1}_{ij} &= \frac{1}{4} \alpha_{2,\pm 1} \bar{\rho} R^5 \left(~
\begin{matrix}
2 \alpha_{2,\pm 1}\cos^2(\omega_{2}t) & \mp \alpha_{2,\pm 1} \sin(2\omega_{2}t) & \mp \frac{8}{15} \sqrt{30\pi}~\cos(\omega_{2}t) \\ 
\mp \alpha_{2,\pm 1} \sin(2\omega_{2}t) & 2 \alpha_{2,\pm 1}\sin^2(\omega_{2}t) & \frac{8}{15} \sqrt{30\pi}~\sin(\omega_{2}t) \\
\mp \frac{8}{15} \sqrt{30\pi}~\cos(\omega_{2}t) & \frac{8}{15} \sqrt{30\pi}~\sin(\omega_{2}t) & 2 \alpha_{2,\pm 1} \\
\end{matrix}
~\right)~, \\
\label{mass_quadrupole_2_2}
\delta I^{2,\pm 2}_{ij} &= \frac{1}{30} \alpha_{2,\pm 2} \bar{\rho} R^5 \left(~
\begin{matrix}
4 \sqrt{30\pi} \cos(\omega_{2}t) + 15 \alpha_{2,\pm 2} & \mp 4 \sqrt{30\pi} \sin(\omega_{2}t) & 0 \\ 
\mp 4 \sqrt{30\pi} \sin(\omega_{2}t) & - 4 \sqrt{30\pi} \cos(\omega_{2}t) + 15 \alpha_{2,\pm 2} & 0 \\
0 & 0 & 0 \\
\end{matrix}
~\right)~,
\end{align}
which have traces equal to $2 \alpha^2_{2,0} \bar{\rho} R^5 \cos^2(\omega_{2}t)$, $\alpha^2_{2,\pm 1} \bar{\rho} R^5$
and $\alpha^2_{2,\pm 2} \bar{\rho} R^5$ for $m = 0, \pm 1, \pm 2$ respectively. The superscripts on $\delta {I}_{ij}$ represent values of $l$ and $m$. As an independent check, we can now take these traces along with the energies in Table~\ref{table:energies} to show that the virial equation in equation~(\ref{virial_equation}) is indeed satisfied by the Kelvin modes for all $m$.

Taking these $\delta {I}_{ij}$, reducing by their trace, differentiating three times with respect to time and substituting into equation~(\ref{GW_luminosity_quadrupole_formula}), we find
\begin{align}
\label{E_GW_Kelvin_mode}
\dot{E}_\text{GW} = \frac{1}{5c^5} \alpha_{2,m}^2 \bar{\rho} \omega_{2}^8 R^{10}~,
\end{align}
for all $m$ and to second order in $\alpha_{2,m}$. To get to this form, we used equation~(\ref{kelvin_mode_frequency_general_l}) which takes a factor of $\bar{\rho}$ along with some other constants and converts them to $\omega_{2}^2$. This is why $\dot{E}_\text{GW}$ is proportional to $\omega_{2}^8$ instead of $\omega_{2}^6$ which would be expected from just the time derivatives alone. 

Our $m=0$ case matches equation~(16) of \citet{chau1967} if one converts from our amplitude parameter, $\alpha_{2,0}$, to theirs. This can be done by equating the radial position of the perturbed surface, which is $R+\bm{\xi}(R, \theta, \phi, t)\cdot \bm{e}_r$ for our work. As far as we are aware, this is the first time the $m \ne 0$ GW luminosities for the Kelvin modes have been written down analytically. For our particular choice of amplitude parameter, the relation for $\dot{E}_\text{GW}$ is the same for all $m$, mirroring the behaviour seen in compressible stars \citep{thorne1969iv}.

\subsection{Rate of change of angular momentum}
\label{subsection: angular momentum considerations}

Moving on, the rate of change of angular momentum taken away by GWs is given by \citep[e.g.][]{misnerThorneWheeler1973book, andersson2019book}
\begin{align}
\label{GW_rate_of_change_angular_momentum_quadrupole_formula}
\dot{J}^i_\text{GW} = \frac{2}{5} \frac{G}{c^5} \varepsilon^{ijk} \left\langle \mathsout{\ddot{I}}_{~j}^{\hspace{1pt}l} \mathsout{\dddot{I}}_{kl} \right\rangle~,
\end{align}
which is positive when GWs are carrying away positive angular momentum. We find only the $z$ component, $\dot{J}^z_\text{GW}$, is present so we set $\dot{J}_\text{GW} = \dot{J}^z_\text{GW}$. This means angular momentum is only being lost in the $z$ direction (for $m \ne 0$). 

We follow the same steps as before where we take $\delta I_{ij}$, which we have already from equations~(\ref{mass_quadrupole_2_0})~--~(\ref{mass_quadrupole_2_2}), reduce by the trace and take an appropriate number of time derivatives to use in the above equation for $\dot{J}_\text{GW}$. After some algebra, we find
\begin{align}
\label{GW_rate_of_change_angular_momentum_Kelvin_mode}
\dot{J}_\text{GW} = - \frac{1}{5c^5} m \alpha_{2,m}^2 \bar{\rho} \omega_{2}^7 R^{10}~.
\end{align}

\subsection{Gravitational wave damping time-scales}
\label{subsection: GW_timescales}

Combining our results for the energy perturbation $\delta E$ (equation~(\ref{eq:delta_E})) and the angular momentum perturbation $\delta J$ (equation~(\ref{eq:delta_J})) we see that 
\begin{equation}
\label{eq:delta_E_delta_J_relation}
\delta E = 2 \omega_{\rm p} \delta J~,
\end{equation}
for $m \ne 0$. If instead we combine our results for the rate at which GW energy is radiated to infinity (equation~(\ref{E_GW_Kelvin_mode})) and the rate at which angular momentum is radiated (equation~(\ref{GW_rate_of_change_angular_momentum_Kelvin_mode})), we see, for $m \ne 0$, that
\begin{equation}
\label{eq:E-dot_delta_J-dot_relation}
\dot E_{\rm GW} = \omega_{\rm p} \dot J_{\rm GW}~.
\end{equation}
This is a well-known relation that holds true for GW emission from any mass distribution that has a well-defined pattern speed, and is an immediate consequence of the standard quadrupole formulae of equations~(\ref{GW_luminosity_quadrupole_formula}) and (\ref{GW_rate_of_change_angular_momentum_quadrupole_formula}). The energy and angular momentum radiated comes at the expense of the energy and angular momentum of the star. 

In the case $m=0$, we have $\delta J=0$ and $\dot J_{\rm GW} = 0$, and the radiated energy clearly comes at the expense of the mode energy. Given that both $\delta E$ and $\dot E_{\rm GW}$ are quadratic in $\alpha_{2,m}$, the mode decays exponentially on a time-scale
\begin{align}
\label{timescale_from_energy_definition}
\tau_\text{E} \equiv \frac{2 \delta E}{\dot{E}_\text{GW}}~. 
\end{align}
Using equations~(\ref{eq:delta_E}) and (\ref{E_GW_Kelvin_mode}), we obtain
\begin{align}
\label{energy_timescale}
\tau_\text{E} = \frac{10c^5}{\omega_2^6 R^5}~.
\end{align}
Note that the $m=0$ result can be compared to equation~(19) of \cite{chau1967} but since his $\delta E$ is a factor of 2 too small (see Section~\ref{subsection: mode energies}), his result for $\tau_\text{E}$ is also a factor of 2 too small. Equation~(\ref{energy_timescale}) is also valid for $m \ne 0$ and in this case, we can check our results against \citet{detweiler1975} to which we find our results are consistent.

For $m \neq 0$, the situation is more interesting. Comparing equations~(\ref{eq:delta_E_delta_J_relation}) and (\ref{eq:E-dot_delta_J-dot_relation}), we see that while $\delta E$ is locked to $\delta J$, and $\dot E_{\rm GW}$ is locked to $\dot J_{\rm GW}$, the constants of proportionality differ by a factor of $2$. This mean that, unlike the $m=0$ case, the response of the star to the GW emission cannot simply be a decrease in mode amplitude.  More explicitly, we can argue as follows:  in a non-rotating star (of given $\bar{\rho}$, $M$ and $R$), the mode itself is parametrised by the single variable $\alpha_{2, m}$, and its energy by $\alpha_{2,m}^2$.  It follows that any other quantity proportional to $\alpha_{2, m}^2$, including $\delta J$, must evolve on the same time-scale as given in equation~(\ref{energy_timescale}).  However, if we were to attempt to compute the angular momentum-based analogue of equation  (\ref{energy_timescale}), we would define 
\begin{align}
\label{timescale_from_angular_momentum_definition}
\tau_\text{J} \equiv \frac{2 \delta J}{\dot{J}_\text{GW}}~,
\end{align}
which leads to
\begin{align}
\label{angular_momentum_timescale}
\tau_\text{J} = \frac{5c^5}{\omega_2^6 R^5}~,
\end{align}
i.e.~there is a factor of $2$ mismatch between the energy-based and angular momentum-based time-scales.  

The natural resolution to this seeming contradiction is to realise that in defining the time-scale of equation~(\ref{timescale_from_angular_momentum_definition}), we have implicitly assumed that all of the angular momentum radiated to infinity (i.e.~all of $\dot{J}_\text{GW}$) comes entirely at the expense of the mode angular momentum (i.e.~comes entirely from $\delta J$).  This suggests that the resolution to this contradiction is to allow for a torque to be exerted on the bulk star (i.e.\ on the background configuration), something which must then be accounted for in the angular momentum balance in order to obtain consistent results.

\section{Canonical mode energies and angular momenta}
\label{section: canonical mode energies and angular momenta}

In the previous section, we showed that for $m \neq 0$, the effect of radiation reaction is to exert a torque on the star, such that it must acquire a non-zero angular velocity. This means one must be careful in allowing for the development of rotation in the radiation reaction calculation, despite our non-rotating initial configuration. The complications and subtleties that arise from rotation in stellar perturbation theory are precisely those captured by the canonical energy/angular momentum formalism of \citet{friedmanschutz1978, friedmanschutz1978b}. We will therefore now proceed to employ this machinery to calculate carefully the response of the star to the emission of the radiation, both in terms of the mode amplitude and the development of rotation.  We will calculate the canonical energy of our Kelvin modes in Section~\ref{subsection: canonical energy}, and the canonical angular momentum in Section~\ref{subsection: canonical angular momentum}. Then, in Section~\ref{section: gravitational wave back-reaction}, we will compute the time derivatives of these quantities, which will allow a calculation of the effect of radiation reaction for $m \neq 0$.

\subsection{Canonical energy}
\label{subsection: canonical energy}

 \citet{friedmanschutz1978, friedmanschutz1978b} noted that, in addition to the physical second order energy perturbation $\delta E$, it is useful to define a \emph{canonical energy} (described below), particularly when dealing with rotating stars.
Although the following equations were originally derived for compressible NSs, we have independently checked from first principles that they remain valid for incompressible NSs.
The first part of this section lays out the general equations, and the second part applies them to the Kelvin modes.


\subsubsection{General equations}

\citet{friedmanschutz1978} noted that the equation of motion can be written in the form
\begin{align}
\label{equation_of_motion_non_dissipative}
A^i_{~j} \ddot{\xi}^j + B^i_{~j} \dot{\xi}^j + C^i_{~j} \xi^j \equiv 0~,
\end{align}
(FSa15) where $A$, $B$ and $C$ are operators that depend on properties the background NS.  For a non-rotating NS, $B^i_{~j}=0$. However, we keep all terms for now as this subsection is general and applies to all NSs. 
The form of $A^i_{~j}$ and $C^i_{~j}$ for our non-rotating incompressible star are given in Section~\ref{subsubsection: application_to_Kelvin_modes_energy_inertial_frame} below. By writing the perturbed equation of motion this way, the Lagrangian density $\mathcal{L}$, can be written as
\begin{align}
\label{lagrangian_general}
\mathcal{L} = \frac{1}{2} \left( \dot{\xi}_i A^i_{~j} \dot{\xi}^j + \dot{\xi}_i B^i_{~j} \xi^j - \xi_i C^i_{~j} \xi^j \right)~,
\end{align}
(FSa35). The canonical energy (in the inertial frame), $E_\text{c}$, is defined as
\begin{align}
\label{canonical_energy_definition}
E_\text{c} \equiv \int_{V} \left( \dot{\xi}^i \frac{\partial \mathcal{L}}{\partial \dot{\xi}^i} - \mathcal{L} \right) dV~,
\end{align}
(FSa44) where $\dot{\xi}^i = \frac{\partial \xi^i}{\partial t}$ is real. 
Substituting the Lagrangian into equation~(\ref{canonical_energy_definition}), we get our first method of calculating $E_\text{c}$ (using real $\xi$)
\begin{align}
\label{canonical_energy_real_xi}
E_\text{c} = \frac{1}{2} \int_{V} \left( \dot{\xi}_i A^i_{~j} \dot{\xi}^j + \xi_i C^i_{~j} \xi^j \right) dV~.
\end{align}

Alternatively, \citet{friedmanschutz1978} found that a ``sympletic structure'' could more neatly describe a perturbed system. One key difference is that this sympletic structure requires complex $\xi$, which, for clarity, we will write as $\tilde\xi$.  From this, \citet{friedmanschutz1978} provided another method to calculate $E_\text{c}$ which we will label $\tilde{E}_\text{c}$, since it will be calculated with $\tilde\xi$. The following equation is true for oscillation modes that carry a $e^{i\omega_{l}t}$ dependence
\begin{align}
\label{canonical_energy_complex_general}
\tilde{E}_\text{c} = \omega_{l} \left[ \text{Re}\{\omega_{l}\} \langle \tilde{\xi}, A \tilde{\xi} \rangle - \frac{1}{2} \langle \tilde{\xi}, i B \tilde{\xi} \rangle \right]~,
\end{align}
(FSa50) where the angled brackets represents a complex inner product, defined as
\begin{align}
\langle \tilde{\xi}, \tilde{\eta} \rangle \equiv \int_{V} (\tilde{\xi}^i)^* \tilde{\eta}_i dV~,
\end{align}
(above FSa36). To quickly summarise, there are two methods to calculate $E_\text{c}$, one using equation~(\ref{canonical_energy_real_xi}) with real $\xi$ and the other using equation~(\ref{canonical_energy_complex_general}) with complex $\xi$. 

\citet{friedmanschutz1978} found a simple relation between the physical second order energy perturbation $\delta E$ and the canonical energy $E_{\rm c}$
\begin{align}
\label{physical_energy_extra_integral}
\delta E = E_\text{c} + \int_{V} \rho v^i \Delta v_i dV~,
\end{align}
(FSa59) where $\rho$ and $v^i$ are the mass density and velocity of the unperturbed NS, and $\Delta v_i$ is the second order (covariant) Lagrangian change to the velocity. 


One can see from equation~(\ref{physical_energy_extra_integral}) that for a static background ($v^i=0$), the integral vanishes and we are left with the physical energy equalling the canonical energy, i.e.~$\delta E = E_\text{c}$. We nevertheless proceed to calculate $E_{\rm c}$ using the formulae given above, as a check on our expression for $\delta E$ in equation~(\ref{eq:delta_E}). The calculation also provides a natural precursor to the calculation of the canonical angular momentum in the next section, which does \emph{not} equal the physical perturbation $\delta J$.


\subsubsection{Application to Kelvin modes}
\label{subsubsection: application_to_Kelvin_modes_energy_inertial_frame}

We will now apply the preceding equations to the Kelvin modes. We will calculate the canonical energy, $E_\text{c}$, which should be the same regardless of whether it is calculated with real or complex $\xi$, and must be equal to $\delta E$ calculated previously (equation~(\ref{eq:delta_E})).

We start with the case of using real $\xi$ to calculate ${E}_\text{c}$, see equation~(\ref{canonical_energy_real_xi}). To do so, we recall the form of the equation of motion
\begin{align}
\rho \frac{\partial \mathbf{v}}{\partial t} + \rho (\mathbf{v} \cdot \bm{\nabla}) \mathbf{v} = - \bm{\nabla} P - \rho \bm{\nabla}\Phi~,
\end{align}
which we perturb with (Eulerian) perturbations, $Q \rightarrow Q_0 + \delta Q$, keeping first order terms only. Then, we subtract the background solution and enforce our assumptions of a static, uniformly-dense, incompressible NS to get
\begin{align}
\bar{\rho} \ddot{\bm{\xi}} + \bm{\nabla} \delta P + \bar{\rho} \bm{\nabla}\delta \Phi = 0~,
\end{align}
where we used $\delta \mathbf{v} = \dot{\bm{\xi}}$. We could have also obtained this from (FSa15) by using the conditions $\nabla_i \xi^i = 0$ and $v^i =0$, alongside a re-expression of $\Phi$ in terms of $P$. Comparing to equation~(\ref{equation_of_motion_non_dissipative}), one immediately finds
\begin{align}
A^i_{~j} &= \bar{\rho} \delta^i_{~j}~, \\
B^i_{~j} &= 0~, \\
C^i_{~j} \xi^j &= \nabla^i \delta P + \bar{\rho} \nabla^i \delta \Phi~,
\end{align}
where $\delta^i_{~j}$ is the Kronecker delta. Note that $B^i_{~j}$ is zero as expected for our static background configuration.

The expression for $A^i_{~j}$ is straightforward and for $C^i_{~j} \xi^j$, we need to know what $\delta P$ and $\delta \Phi$ are, which were presented earlier in equations~(\ref{delta_P}) and (\ref{delta_Phi}). Using this along with $\xi$ from equation~(\ref{real_xi_general_l}), equation~(\ref{canonical_energy_real_xi}) in integral form becomes
\begin{align}
\label{final_canonical_energy_real}
E_\text{c} = \frac{1}{2} \alpha_{lm}^2 \bar{\rho} \omega_{l}^2 \int_{V} r^{2l-2}  \left[ 
l^2  Y_{lm}^2(\theta, 0) 
+  \left(\frac{d Y_{lm}(\theta, 0)}{d\theta}\right)^2 
+ \frac{m^2}{\sin^2\theta} Y_{lm}^2(\theta, 0)
\right] dV~,
\end{align}
for all $l$ and $m$. When we evaluate this integral for $l=2$ Kelvin modes, we find that the canonical energy is given by
\begin{align}
\label{canonical_energy_Kelvin_mode}
\delta E = E_\text{c} = \alpha^2_{2,m} \bar{\rho } \omega _2^2 R^5~,
\end{align}
i.e.~we have agreement with the physical second order energy perturbation $\delta E$ computed in Section~\ref{subsection: mode energies}.

Alternatively, we can use complex $\tilde\xi$ to find $\tilde E_\text{c}$, see equation~(\ref{canonical_energy_complex_general}). The form of the complex solution is basically given by equation~(\ref{real_xi_general_l_Re}) above, where we now remove the $\Re$ operator and also, importantly, insert an additional normalisation factor $N$ 
\begin{align}
\label{complex_xi_general_l}
\tilde{\bm{\xi}} (r, \theta, \phi, t) &= N \frac{\alpha_{lm}}{R^{l-2}} \bm{\nabla} [r^l Y_{lm}(\theta, \phi)] e^{i\omega_{l}t}~, \\
\label{complex_xi_general_l_vector_notation}
\tilde{\bm{\xi}} (r, \theta, \phi, t) &= N \alpha_{lm} \frac{r^{l-1}}{R^{l-2}}  
\left[ l Y_{lm}(\theta, 0) \bm{e}_r 
+  \frac{d Y_{lm}(\theta, 0)}{d\theta}\bm{e}_\theta 
+  \frac{im }{\sin\theta} Y_{lm}(\theta, 0) \bm{e}_\phi\right]   e^{i(m\phi + \omega_{l} t)}~.
\end{align}
Using $A^i_{~j}$ (and $B^i_{~j}= 0$) in equation~(\ref{canonical_energy_complex_general}), along with $\tilde\xi$ from equation~(\ref{complex_xi_general_l_vector_notation}), we find that the canonical energy of the Kelvin modes in integral form is 
\begin{align}
\label{final_canonical_energy_complex}
\tilde{E}_\text{c} = N^2 \alpha_{lm}^2 \bar{\rho} \omega_{l}^2 \int_{V} r^{2l-2} 
\left[ 
l^2 Y_{lm}^2(\theta, 0)  
+  \left(\frac{d Y_{lm}(\theta, 0)}{d\theta}\right)^2 
+  \frac{m^2}{\sin^2\theta} Y_{lm}^2(\theta, 0)  
\right] dV~.
\end{align}

Since the canonical energy must be the same regardless of whether $\xi$ is real or complex, i.e.~$E_\text{c} = \tilde{E}_\text{c}$, we find from comparing equations~(\ref{final_canonical_energy_complex}) and (\ref{final_canonical_energy_real}) that 
\begin{align}
N^2 = \frac{1}{2} ~~ \rightarrow ~~ N = \frac{1}{\sqrt{2}}~,
\end{align}
for all $l$ and $m$. With this considered, equations~(\ref{final_canonical_energy_complex}) and (\ref{final_canonical_energy_real}) now give the same canonical energy.

This normalisation holds for all modes and not just the Kelvin modes -- the proof is shown in Appendix~\ref{appendix: real and complex mode eigenfunctions}. This simple result is useful and necessary whenever there is a mixture of real and complex $\xi$ within an analysis. Since the equation of motion is linear in the perturbation, there is no right or wrong relative normalisation for $\xi$ as such, but one \textit{must} enforce a relative normalisation if one wants to use both real and complex formulae in the same calculation. Without normalising, the calculation of the canonical energy with complex $\xi$ would be twice the value obtained when the calculation is done entirely with real expressions. This clearly cannot be correct. To resolve this, one must put a factor of $\frac{1}{\sqrt{2}}$ in front of each complex $\xi$ (specifically, a factor of $\frac{1}{\sqrt{2}}$ for every $\alpha_{lm}$). This can be understood intuitively if one thinks of complex $\xi$ as comprising of an equal amount of ``power'' in its real and imaginary parts. Thus, if we only consider the real part, we only get half the power, and hence $\frac{1}{\sqrt{2}}$ the amplitude.

\subsection{Canonical angular momentum} 
\label{subsection: canonical angular momentum}

\subsubsection{General equations}


We again begin by giving the general equations, to be applied to the Kelvin modes in Section~\ref{subsubsection: application to Kelvin modes angular momentum}. The canonical angular momentum is defined as
\begin{align}
\label{canonical_angular_momentum_definition}
J_\text{c} \equiv - \int_{V} \frac{\partial \xi^i}{\partial \phi} \frac{\partial \mathcal{L}}{\partial \dot{\xi}^i} dV~,
\end{align}
(FSa47) where $\phi$ is the azimuthal angle and $\xi$ is real. Substituting in the Lagrangian from equation~(\ref{lagrangian_general}) gives
\begin{align}
\label{canonical_angular_momentum_real_xi_definition}
J_\text{c} = - \int_{V} \frac{\partial \xi^i}{\partial \phi} \left(A_{ij} \dot{\xi}^j + \frac{1}{2} B_{ij} \xi^j \right) dV~,
\end{align}
which will be the calculation using real $\xi$. 

Alternatively, the sympletic structure can be used to find $J_\text{c}$ but requires $\xi$ to be complex and have a $e^{i(m\phi+\omega_{l}t)}$ dependence. \citet{friedmanschutz1978} found this to be
\begin{align}
\label{canonical_angular_momentum_complex_xi_definition}
\tilde{J}_\text{c} = - m \left[ \text{Re}\{\omega_{l}\} \langle \tilde{\xi}, A \tilde{\xi} \rangle - \frac{1}{2} \langle \tilde{\xi}, i B \tilde{\xi} \rangle \right]~,
\end{align}
(FSa51) for complex $\xi$. Comparing this to equation~(\ref{canonical_energy_complex_general}), one finds, for $m\ne0$, that $E_\text{c}$ and $J_\text{c}$ are related by the pattern speed
\begin{align}
\label{E_c_J_c_relation}
E_\text{c} = - \frac{\omega_{l}}{m} J_\text{c} = \omega_\text{p} J_\text{c}~,
\end{align}
(FSa52) which offers a quick alternative method to calculate $ J_\text{c}$ if one already has $E_\text{c}$. 

Note that whereas, for a non-rotating star, the physical and canonical energy of the perturbations were the same, this is not the case for the physical and canonical angular momenta
\begin{align}
\label{physical_angular_momentum_extra_integral}
\delta J = J_\text{c} + \int_{V} \rho \phi^i \Delta v_i dV~,
\end{align}
(FSa61). This point will be of importance when calculating the effect of radiation reaction.


\subsubsection{Application to Kelvin modes}
\label{subsubsection: application to Kelvin modes angular momentum}

Once again, we specialise to the $l=2$ Kelvin modes.
Since we already have $E_\text{c}$ from equation~(\ref{canonical_energy_Kelvin_mode}), the simplest method would be to utilise equation~(\ref{E_c_J_c_relation}) which gives
\begin{align}
\label{canonical_angular_momentum_Kelvin_mode}
J_\text{c} = - m \alpha^2_{2,m} \bar{\rho } \omega _2 R^5~.
\end{align}
We verified that one obtains the same value when using equation~(\ref{canonical_angular_momentum_real_xi_definition}) with real $\xi$ or equation~(\ref{canonical_angular_momentum_complex_xi_definition}) with complex $\xi$ (whilst ensuring to account for the extra factor of $\frac{1}{\sqrt{2}}$ for every $\alpha_{2,m}$).


Note that this does indeed differ from the physical angular momentum of the Kelvin modes as given by equation~(\ref{physical_angular_momentum_extra_integral}), corresponding to non-vanishing of the last term in equation~(\ref{physical_angular_momentum_extra_integral})
\begin{align}
\label{extra_term}
\delta J - J_{\rm c} = \int_{V} \rho \phi^i \Delta v_i dV = \frac{1}{2} m \alpha^2_{2,m} \bar{\rho } \omega _2 R^5~.
\end{align}
Related to this, comparing equations~(\ref{eq:delta_J}) and (\ref{canonical_angular_momentum_Kelvin_mode}) we see that
\begin{equation}
\label{eq:J_c-delta_J}
J_{\rm c} = 2 \delta J~.
\end{equation}


\section{Gravitational wave back-reaction}
\label{section: gravitational wave back-reaction}

\subsection{Mode damping time-scale}
\label{subsection: using the canonical energy in the rotating frame}

\subsubsection{General equations}

The aim here is to highlight a calculation, based on the work of \citet{friedmanschutz1978, friedmanschutz1978b}, that can determine the damping time-scale of modes on any NS. 
After the general calculation, we specialise to the Kelvin modes which will be covered in Section~\ref{subsection: application_to_Kelvin_modes_rotating_frame}. 
The calculational method described here is similar to one done by \citet{ipserLindblom1991}, but we stick rigidly to the formalism of Friedman \& Schutz.


 \citet{friedmanschutz1978} define the \emph{canonical energy in the rotating frame}, $E_\text{c,R}$, by
\begin{align}
\label{connecting_E_cR_E_c_delta_E}
E_\text{c,R} = E_\text{c} - \Omega J_\text{c} = \delta E - \Omega \delta J~,
\end{align}
(FSa62) where $\Omega$ is the angular frequency of the rigidly rotating NS. 
Note that $E_\text{c,R} = E_\text{c} = \delta E$ for non-rotating NSs.

One key reason why we introduced $E_\text{c,R}$ is because the time derivative of $E_\text{c,R}$ is easily calculable, thus making the physical mode damping time-scale also easily calculable. To do this, we first need the perturbed form of the equation of motion of a dissipative system, which is given by
\begin{align}
\label{general_equation_of_motion_dissipative}
A^i_{~j} \partial_t^2{\xi}^j + B^i_{~j} \partial_t{\xi}^j + C^i_{~j} \xi^j = F^i~,
\end{align}
(FSb46) where (as always) $\partial_t$ represents a time derivative in the inertial frame and crucially, $F^i$ is some dissipative force (per unit volume) acting on the system. From this, one can calculate the time derivative of $E_\text{c,R}$ with
\begin{align}
\label{rate_of_change_E_c_R_definition}
\frac{d}{dt} E_\text{c,R} = \langle \dot{\tilde{\xi}}, \tilde{F} \rangle~,
\end{align}
(FSb49).  (Formally, the dot here represents a time derivative evaluated in the rotating frame, i.e.~$\dot f = (\partial_t + \Omega \partial_\phi)f$ (FSb42), but this distinction is not important for the non-rotating stars considered here.)  This time derivative of $E_\text{c,R}$ is negative when the system is losing energy. It is also worth remembering that $\dot{\xi}$ and $F$ must be complex in this equation (but the result of taking the inner product will be real; see below).

The mode damping time-scale is then given by
\begin{align}
\label{physical_decay_timescale}
\tau_\text{phys} \equiv - \frac{2 E_\text{c,R}}{\frac{d}{dt} E_\text{c,R}}~,
\end{align}
where we have inserted the minus sign to adhere to the convention that damping times are positive for the stable modes considered here.  Unlike the earlier definitions of $\tau_\text{E}$ (equation~(\ref{timescale_from_energy_definition})) and $\tau_\text{J}$ (equation~(\ref{timescale_from_angular_momentum_definition})), $\tau_\text{phys}$ has the same quantities in the numerator and denominator, and so is clearly the correct expression for the mode damping time-scale. More explicitly, the numerators of $\tau_\text{E}$ and $\tau_\text{J}$ were the physical mode energy and angular momentum whereas the denominators were the rate of emission of GW energy and angular momentum.

There are a few more quantities that can be calculated with this formalism, where the equation of motion explicitly includes a dissipative term. For instance, the rate of change in the canonical energy and angular momentum can be calculated with
\begin{align}
\label{rate_change_canonical_energy}
\frac{d}{dt} E_\text{c} &= \text{Re} \langle \partial_t\tilde{\xi}, \tilde{F} \rangle~, \\
\label{rate_change_canonical_angular_momentum}
\frac{d}{dt} J_\text{c} &= - \text{Re} \langle \partial_\phi\tilde{\xi}, \tilde{F} \rangle~,
\end{align}
(FSb70 \& FSb71), where we have complex quantities on the right-hand side. Both are negative when canonical energy or positive angular momentum is being lost from the system. From this, \cite{friedmanschutz1978b} found that the rate of change of $E_\text{c}$ and $J_\text{c}$ are related by the pattern speed
\begin{align}
\frac{d}{dt} E_\text{c} = \omega_\text{p} \frac{d}{dt} J_\text{c}~,
\end{align}
(below FSb71) much like the $\dot{E}_\text{GW}$ and $\dot{J}_\text{GW}$ in equation~(\ref{eq:E-dot_delta_J-dot_relation}).

\subsubsection{Application to Kelvin modes}
\label{subsection: application_to_Kelvin_modes_rotating_frame}

Like in Section~\ref{subsubsection: application_to_Kelvin_modes_energy_inertial_frame}, we write down the equation of motion for our system, perturb it, keep first order terms, subtract the background solution and apply our assumptions. What is different now is that we allow for a dissipative force, GW emission in this example, in the equation of motion. For GW emission, the dissipative force is given by
\begin{align}
\label{gravitational_radiation_reaction_force_definition}
F^a = - \bar{\rho} \nabla^a \Phi_\text{RR}~,
\end{align}
where $\Phi_\text{RR}$ is the Burke-Thorne GW radiation reaction potential 
\begin{align}
\label{radiation_reaction_potential_definition}
\Phi_\text{RR} = \frac{1}{5} \frac{G}{c^5} x^i x^j \frac{d^5}{dt^5} \mathsout{I}_{ij}~,
\end{align}
\citep{burke1969, burke1971, thorne1969iv, misnerThorneWheeler1973book}. This equates to having a perturbed equation of motion (\ref{general_equation_of_motion_dissipative}) of the form
\begin{align}
\label{equation_of_motion_dissipation}
\bar{\rho} \ddot{\xi}^a + \nabla^a \delta P + \bar{\rho} \nabla^a\delta \Phi = - \frac{1}{5} \frac{G}{c^5} \bar{\rho} \nabla^a x^i x^j \frac{d^5}{dt^5} \mathsout{I}_{ij}~,
\end{align}
where $a$ labels a spherical basis and $i, j$ labels a Cartesian basis. When written like this, it is clear we must use real $\xi$ in the preceding equations, including for $F^a$ in equation~(\ref{gravitational_radiation_reaction_force_definition}). However, we saw earlier that we need complex $F^a$ in equations~(\ref{rate_of_change_E_c_R_definition}), (\ref{rate_change_canonical_energy}) and (\ref{rate_change_canonical_angular_momentum}). This reiterates our need to connect real and complex expressions, and as shown in Section~\ref{subsubsection: application_to_Kelvin_modes_energy_inertial_frame} and Appendix~\ref{appendix: real and complex mode eigenfunctions}, we can do so by introducing a factor of $\frac{1}{\sqrt{2}}$ for every $\alpha_{lm}$ when going from real to complex.

To find $\tau_\text{phys}$, we first need to find an expression for $E_\text{c,R}$, but this is trivial for a non-rotating NS. For a non-rotating NS, $E_\text{c,R}$ is simply equal to $E_\text{c}$ so we read directly from equation~(\ref{canonical_energy_Kelvin_mode})
\begin{align}
\label{canonical_energy_rotating_frame_Kelvin_mode}
E_\text{c,R} = E_\text{c} = \alpha^2_{2,m} \bar{\rho } \omega _2^2 R^5~,
\end{align}
for all $m$.

Next is to find $\frac{d}{dt} E_\text{c,R}$ so we now focus on $F^a$. For the $l=2$ Kelvin modes, we get $\mathsout{I}_{ij}$ by manipulating $\delta I_{ij}$ from equations~(\ref{mass_quadrupole_2_0})~--~(\ref{mass_quadrupole_2_2}). We follow the same steps as before where we reduce by the trace (to get $\delta \mathsout{I}_{ij}$) and since $\mathsout{\dot{I}}_{ij} = \delta \mathsout{\dot{I}}_{ij}$ (which also holds for all higher time derivatives), we differentiate $\delta \mathsout{I}_{ij}$ five times and substitute into equation~(\ref{radiation_reaction_potential_definition}) to find
\begin{align}
\label{gravitational_radiation_reaction_potential_all_m}
\Phi_\text{RR} \approx 
\begin{cases}
- \frac{2\sqrt{5 \pi}}{75} \frac{G}{c^5} \alpha _{2,0} \bar{\rho } \omega _2^5 R^5 r^2 (1+ 3\cos 2 \theta) \sin (\omega _2 t) & ~~\text{for}~ m = 0~, \\
\pm \frac{2\sqrt{30 \pi } }{75} \frac{G}{c^5} \alpha _{2,\pm 1} \bar{\rho } \omega _2^5 R^5 r^2 \sin 2 \theta \sin (\pm \phi + \omega _2 t) & ~~\text{for}~ m = \pm 1~, \\
- \frac{2\sqrt{30 \pi } }{75} \frac{G}{c^5} \alpha _{2,\pm 2} \bar{\rho } \omega _2^5 R^5 r^2 \sin ^2\theta \sin (\pm 2 \phi + \omega _2 t) & ~~\text{for}~ m = \pm 2~,
\end{cases} 
\end{align}
to first order in $\alpha_{2,m}$. Next, we use the GW radiation reaction potential to calculate the GW radiation reaction force using equation~(\ref{gravitational_radiation_reaction_force_definition})
\begin{align}
F^a &= \frac{4\sqrt{5 \pi}}{75} \frac{G}{c^5} \alpha _{2,0} \bar{\rho }^2 \omega _2^5 R^5 r 
\begin{pmatrix} 
(1+ 3\cos 2 \theta) \sin (\omega _2 t) \\
-3 \sin 2 \theta \sin (\omega _2 t) \\ 
0
\end{pmatrix}\hspace{0.91cm}\text{for}~ m = 0~, \\
F^a &= - \frac{4\sqrt{30 \pi } }{75} \frac{G}{c^5} \alpha _{2,\pm 1} \bar{\rho }^2 \omega _2^5 R^5 r
\begin{pmatrix} 
\pm \sin 2 \theta \sin (\pm \phi + \omega _2 t) \\
\pm \cos 2 \theta \sin (\pm \phi + \omega _2 t ) \\ 
\cos \theta \cos (\pm \phi + \omega _2 t )
\end{pmatrix}\hspace{0.33cm}\text{for}~ m = \pm 1~, \\
F^a &= \frac{2\sqrt{30 \pi } }{75} \frac{G}{c^5} \alpha _{2,\pm 2} \bar{\rho}^2 \omega _2^5 R^5 r 
\begin{pmatrix} 
2 \sin ^2\theta \sin (\pm 2 \phi + \omega _2 t ) \\
\sin 2 \theta \sin (\pm 2 \phi + \omega _2 t ) \\ 
\pm 2 \sin \theta \cos (\pm 2 \phi + \omega _2 t )
\end{pmatrix}\hspace{0.37cm}\text{for}~ m = \pm 2~,
\end{align}
which is purely real and comes from using real $\xi$. Here, $F^a$ is written in terms of spherical basis vectors ($\bm{e}_r$, $\bm{e}_\theta$, $\bm{e}_\phi$). This is almost what we want, but equation~(\ref{rate_of_change_E_c_R_definition}) is only valid for complex $F^a$. To ``complexify'' our real expression, we use the following
\begin{align}
\cos (m \phi \pm \omega _2 t)~ &\rightarrow~ \frac{1}{\sqrt{2}} e^{i (m \phi \pm \omega _2 t)} \\
\sin (m \phi \pm \omega _2 t)~ &\rightarrow~ - \frac{1}{\sqrt{2}} i e^{i (m \phi \pm \omega _2 t)}
\end{align}
where the factor of $\frac{1}{\sqrt{2}}$ comes from Section~\ref{subsubsection: application_to_Kelvin_modes_energy_inertial_frame} (or Appendix~\ref{appendix: real and complex mode eigenfunctions}) and from the fact that $F^a$ depends linearly on $\alpha_{2,m}$. This then gives the complex expression for $F^a$ which is
\begin{align}
\tilde F^a &= - \frac{1}{\sqrt{2}} \frac{4\sqrt{5 \pi}}{75} \frac{G}{c^5} \alpha _{2,0} \bar{\rho }^2 \omega _2^5 R^5 r e^{i \omega _2 t}
\begin{pmatrix} 
i (1+ 3\cos 2 \theta) \\
- 3 i \sin 2 \theta \\ 
0
\end{pmatrix}~~~~~~~~~\text{for}~ m = 0~, \\
\tilde F^a &= - \frac{1}{\sqrt{2}} \frac{4\sqrt{30 \pi } }{75} \frac{G}{c^5} \alpha _{2,\pm 1} \bar{\rho }^2 \omega _2^5 R^5 r e^{i(\pm \phi + \omega _2 t )}
\begin{pmatrix} 
\mp i \sin 2 \theta \\
\mp i \cos 2 \theta \\ 
\cos \theta 
\end{pmatrix}\hspace{0.35cm}\text{for}~ m = \pm 1~, \\
\tilde F^a &= - \frac{1}{\sqrt{2}} \frac{2\sqrt{30 \pi } }{75} \frac{G}{c^5} \alpha _{2,\pm 2} \bar{\rho}^2 \omega _2^5 R^5 r e^{i(\pm 2 \phi + \omega _2 t )}
\begin{pmatrix} 
2 i \sin ^2\theta \\
i \sin 2 \theta \\ 
\mp 2 \sin \theta 
\end{pmatrix}~~~~\text{for}~ m = \pm 2~,
\end{align}
which can now be used in equations~(\ref{rate_of_change_E_c_R_definition}) with complex $\dot{\xi}$, which is the time derivative of equation~(\ref{complex_xi_general_l_vector_notation}) (with $N = \frac{1}{\sqrt{2}}$). Finally, this gives the (purely real) result
\begin{align}
\label{rate_of_change_E_c_R_Kelvin_mode}
\frac{d}{dt}E_\text{c,R} = - \frac{1}{5c^5} \alpha_{2,m}^2 \bar{\rho} \omega_{2}^8 R^{10}~,
\end{align}
for all $m$, where we have also used equation~(\ref{kelvin_mode_frequency_general_l}) whilst setting $l=2$. This expression for the rate of change of $E_\text{c,R}$ is accurate to leading (second) order in $\alpha_{2,m}$ and for the $m=0$ case, had to be time-averaged. Note that this is equal to -$\dot{E}_\text{GW}$ which was calculated with real $\xi$, i.e.~equation~(\ref{E_GW_Kelvin_mode}). It was not obvious to us from the outset that this would be the case, hence the need for this calculation.

Finally, using equation~(\ref{physical_decay_timescale}), we find
\begin{align}
\tau_\text{phys} = \frac{10c^5}{\omega_2^6 R^5}~,
\end{align}
for all $m$. This is the physical damping time-scale for the Kelvin modes if GW emission is the only dissipative mechanism. 
The $m=0$ damping time-scale calculated here agrees with \citet{chau1967} if one accounts for the factor of 2 that is missing from Chau's expression for his mode energy, see the discussion at the end of Section~\ref{subsection: mode energies} for details. Our results also agree with the $m \ne 0$ Kelvin mode damping time-scales found by \citet{detweiler1975} based on the nearly Newtonian limit of the general relativistic formalism of \cite{thorne1969iv}. For a typical $f$-mode frequency of $f_\text{mode} = 2~\text{kHz}$ (where $\omega_{2} = 2 \pi f_\text{mode}$), the physical damping time-scale is $\tau_\text{phys} \approx 0.06~\text{s}$.

Whilst we have $F^a$ in complex form, we can, for completeness, calculate $\frac{d}{dt} E_\text{c}$ and $\frac{d}{dt} J_\text{c}$ using equations~(\ref{rate_change_canonical_energy}) and (\ref{rate_change_canonical_angular_momentum}). Doing this, one finds
\begin{align}
\frac{d}{dt}E_\text{c} &= - \frac{1}{5c^5} \alpha_{2,m}^2 \bar{\rho} \omega_{2}^8 R^{10} = - \dot{E}_\text{GW}~, \\
\label{rate_change_J_c}
\frac{d}{dt}J_\text{c} &= \frac{1}{5c^5} m \alpha_{2,m}^2 \bar{\rho} \omega_{2}^7 R^{10} = - \dot{J}_\text{GW}~.
\end{align}
The energy result agrees with expectations since the inertial frame and rotating frame of a non-rotating NS are the same, so we would expect $\frac{d}{dt}E_\text{c} = \frac{d}{dt}E_\text{c,R}$.  This result of equation~(\ref{rate_change_J_c}) then follows, to maintain consistency between the relations $\dot E_\text{GW} = \omega_\text{p} \dot J_\text{GW}$ (equation~(\ref{eq:E-dot_delta_J-dot_relation})) and $\dot E_\text{c} = \omega_\text{p} \dot J_\text{c}$ (equation (\ref{E_c_J_c_relation})).

\subsection{Torque exerted on the star}
\label{subsection: torque_exerted_on_star}


In Section~\ref{subsection: GW_timescales}, we argued that, for $m \ne 0$, the effect of radiation reaction could not simply be to damp the mode -- the bulk angular momentum of the star must change too. We are now in a position to compute this change. To do so, we turn to the conservation of angular momentum. We will imagine adding by hand an $m \neq 0$ Kelvin mode to an initially non-rotating star, and calculate the (non-zero) value of $\dot\Omega$ that the radiation reaction induces.
Formally, once the star rotates, the mode eigenfunctions would change too, which means the mode energies and angular momenta would need to be calculated again, but in the slow rotation approximation this will be a higher order correction.


For such a set-up, we follow the logic of the $r$-mode analysis of \citet{owenetal1998} and write the total angular momentum of the system as
\begin{align}
J \equiv I \Omega +\delta J~,
\end{align}
where $I$ is the NS's moment of inertia and $I\Omega$ is the angular momentum of the bulk of the NS. (Note that in \citet{owenetal1998} they used $J_\text{c}$ instead of $\delta J$; it is not clear to us why). Differentiating with respect to time, we get a torque-balance equation
\begin{align}
\dot{J} = I \dot{\Omega} +\delta \dot{J}~,
\end{align}
where we have discarded the $\dot{I}$ term since the rotation corrections to $I$ will be of higher order. By conservation of angular momentum, $\dot{J} = - \dot{J}_\text{GW}$ for our system, so that
\begin{align}
\label{torque_balance}
-\dot J_{\rm GW} = I \dot{\Omega} +\delta \dot{J}~.
\end{align}

There are two (completely equivalent) ways to find $\delta \dot{J}$. The first is to say $\delta \dot{J} = - \frac{2\delta J}{\tau_\text{phys}}$ (since the mode exponentially decays over the physical time-scale) and then substitute in $\delta J$ and $\tau_\text{phys}$. Or, more straightforwardly, we can differentiate the relation $J_{\rm c} = 2\delta J$ of equation~(\ref{eq:J_c-delta_J}) to give
\begin{align}
\dot J_{\rm c} = 2 \delta \dot J~,
\end{align}
which can be combined with the relation $\dot J_{\rm c} = - \dot J_{\rm GW}$ of equation~(\ref{rate_change_J_c}) to give
\begin{align}
\label{eq:delta_J_dot-J_GW_dot}
\delta \dot{J} = - \frac{1}{2} \dot{J}_\text{GW}~.
\end{align}
We can use this to eliminate $\delta \dot J$ from equation~(\ref{torque_balance}) to give
\begin{align}
\label{back_reaction}
I \dot{\Omega} = - \frac{1}{2} \dot{J}_\text{GW}~.
\end{align}
Inserting the explicit form of $\dot{J}_\text{GW}$ from equation~(\ref{GW_rate_of_change_angular_momentum_Kelvin_mode}) we obtain
\begin{align}
\label{back_reaction_2}
I \dot{\Omega} = \frac{1}{10c^5} m \alpha_{2,m}^2 \bar{\rho} \omega_{2}^7 R^{10}~.
\end{align}
This shows that if positive angular momentum is being lost from a prograde ($m<0$) mode, then we would have $\dot{\Omega} < 0$ which rotates the initially static NS in the retrograde direction. For a retrograde mode ($m>0$), we have $\dot{\Omega} > 0$, so the NS rotates in the prograde direction. 

To gain further insight, one can instead use equation~({\ref{eq:delta_J_dot-J_GW_dot}) to eliminate $\dot{J}_\text{GW}$ from equation~(\ref{torque_balance}) to give
\begin{align}
\delta \dot J = I \dot\Omega~,
\end{align}
which can be integrated with respect to time
\begin{align}
\int \delta \dot J \, dt = \int I \dot\Omega \, dt ~.
\end{align}
If we imagine that at $t=0$ we deposit angular momentum $\delta J(t=0)$ into a mode on a non-rotating ($\Omega(t=0)=0$) star, and integrate to $t = \infty$, so that the mode has decayed away, we obtain
\begin{align}
\delta J(t=\infty) - \delta J(t=0) = I \Omega(t=\infty) - I \Omega(t=0)~, 
\end{align}
so that
\begin{align}
\label{eq:delta_Omega-delta_J}
I \delta \Omega = - \delta J(t=0)~, 
\end{align}
where $\delta \Omega$ is the change in angular velocity of the star. 

Similarly, we can re-arrange equation~(\ref{back_reaction}) to give
\begin{align}
\dot{J}_\text{GW} = -2 I \dot{\Omega}~,
\end{align}
which integrates to give
\begin{align}
\delta {J}_\text{GW} = -2 I \delta \Omega~,
\end{align}
where $\delta J_{\rm GW}$ is the total angular momentum radiated as GWs. Substituting for $I \delta \Omega$ using equation~(\ref{eq:delta_Omega-delta_J}) we obtain
\begin{align}
\delta J_{\rm GW} = 2 \delta J(t=0)~.
\end{align}
We therefore see that if one deposits an angular momentum $\delta J(t=0)$ in an $m \neq 0$ Kelvin mode of a non-rotating star, a total angular momentum of $2 \delta J(t=0)$ is ultimately radiated to infinity, while the star itself ``recoils'' by acquiring a rotational angular momentum of $- \delta J(t=0)$. In a sense, the GW emission not only acts to reduce the angular momentum in the mode, but actually ``over-extracts'' what is available, so to compensate, the NS rotates in the opposite direction of the mode to ensure angular momentum is conserved. As far as we know, this is the first time this GW back-reaction has been reported.

An alternative and very direct way of obtaining equation~(\ref{eq:delta_Omega-delta_J}) is as follows.  By angular momentum conservation, the angular momentum left in the star after the mode has dissipated must be equal to $\delta J(t=0) - \delta J_\text{GW}$, as $\delta J(t=0)$ is the initially added angular momentum, and  $\delta J_\text{GW}$ is the total angular momentum radiated away.  But from equation~(\ref{rate_change_J_c}) we see that $\dot J_\text{GW} = - \dot J_\text{c}$, integration of which gives us $\delta J_{\rm GW} = J_\text{c}(t=0)$.  It follows that the final angular momentum of the star is $\delta J(t=0) - \delta J_\text{c}(t=0)$.  But from equation~(\ref{eq:J_c-delta_J}) we have $\delta J_\text{c}(t=0) = 2 \delta J(t=0)$, from which we see that $\delta J(t=0) - \delta J_\text{c}(t=0) = - \delta J(t=0)$, consistent with equation (\ref{eq:delta_Omega-delta_J}).  We are indebted to Andrey Chugunov for suggesting this alternative argument to us.


\section{Summary and implications}
\label{section: conclusions}

In this paper, we applied the \citet{friedmanschutz1978, friedmanschutz1978b} formalism to Kelvin modes, the incompressible manifestation of the $f$-modes, to compute the effect of gravitational radiation reaction on the modes and  the star. We focused on the simplest case of an initially non-rotating, uniformly-dense and incompressible NS. This allowed our calculations to be analytic. 

We used results from \citet{friedmanschutz1978} to show that while the mode energy and angular momentum perturbations were locked together by the relation $\delta E = 2\omega_{\rm p} \delta J$, the rates at which energy and angular momenta were radiated were related by $\dot E_{\rm GW} = \omega_{\rm p} \dot J_{\rm GW}$, where $\omega_{\rm p}$ is the mode pattern speed. This difference in proportionality factors implied that the effect of the radiation reaction could not simply be to damp the mode, as one would have intuitively expected; rather, the star must acquire a bulk angular momentum, i.e.\ is set into rotation.

We used results from \citet{friedmanschutz1978, friedmanschutz1978b} to solve this radiation reaction problem. Specifically, we computed the canonical energy $E_{\rm c}$ and canonical angular momentum $J_{\rm c}$. As described in Friedman \& Schutz, these quantities are locked together in the form $E_{\rm c} = \omega_{\rm p} J_{\rm c}$, while the corresponding time evolution is 
$\dot E_{\rm c} = \omega_{\rm p} \dot J_{\rm c}$ under the influence of radiation reaction. For non-rotating stars, $E_{\rm c} = \delta E$, but $J_{\rm c} = 2 \delta J$. It is this mismatch between the canonical and physical angular momentum perturbations that is key in computing the radiation reaction.

Performing the necessary calculations explicitly, we found that if one, by hand, deposits a mode of angular momentum $\delta J$ into a non-rotating star, a total angular momentum of $2\delta J$ is radiated to infinity, while the star acquires a rotational angular momentum of $-\delta J$. This was an unexpected result. As far as we are aware, this is the first time such calculations have been carried out.  \citet{chau1967} considered only the $m = 0$ case, where this unexpected effect does not occur (and we claim made a factor of $2$ error in his damping time-scale). \citet{detweiler1975} did consider the $m\ne0$ case, and the damping time-scales calculated here agreed with his results, but his calculations stopped short of the full radiation reaction problem which considers the effect on angular momentum. This was the main problem addressed in this paper.

The results given here are of potential interest for modelling the evolution of spinning NSs. One could imagine a spinning-down NS undergoing some sort of impulsive event that excites such a mode. Our results then give a slow-rotation approximation description of how the angular momentum deposited in the mode is radiated away. It was precisely this scenario that first aroused our interest in this problem, and we will explore this further in a separate publication.

\section*{Acknowledgements}

The authors would like to thank Nils Andersson, Fabian Gittins, and the other members of the Southampton gravity group, and also Pantelis Pnigouras, for stimulating discussions during the course of this work. We also thank Andrey Chugunov for suggesting an alternative argument for the gravitational wave back-reaction which was provided at the end of Section~6.2. GY acknowledges support from the EPSRC via grant number EP/N509747/1. DIJ acknowledges support from the STFC via grant number ST/R00045X/1.

\section*{Data Availability}

This article did not use any data.



\bibliographystyle{mnras}
\bibliography{references} 




\appendix

\section{Gravitational wave back-reaction from masses connected by springs}
\label{appendix: toy model}

In this appendix, we show the existence of a GW back-reaction on a simple mass-spring system, a result that arises from the conservation of angular momentum. What we mean by a GW back-reaction is the (unexpected) bulk rotation of a rigid system in the direction opposite to whatever is causing GWs (and hence angular momentum) to be emitted. For instance, in the main text, we found the emission of GWs from a prograde $f$-mode caused the NS to rotate in the retrograde direction. The mass-spring problem below is analogous to the NS system in the main text, however, there is a slight difference which will be highlighted at the appropriate time. Nevertheless, we are still able to show the presence of a GW back-reaction which gives us confidence in the (similar) result found for NSs.

Our model consists of $N$ oscillators. Each oscillator is made up of two identical masses $M$ joined by a light (i.e.~massless) spring of natural length $2R$, with the oscillator having a natural oscillation frequency of $\omega$. We place these oscillators so as to form a ring of radius $R$ of equally spaced masses in the $z=0$ plane, with the centre point of each oscillator at the origin, so that the angle between each oscillator and the next is $\pi/N$. Figure~\ref{fig: oscillators} shows a diagram of the system.

\begin{figure}

\centering
\begin{circuitikz}
	\tikzmath{\x = 0.2; \diag1 = 2*1.7071; \diag2 = 2*0.2929;} 
	
	\tikzstyle{spring}=[thick,decorate,decoration={coil,pre length=0.2cm,post
		length=0.2cm,segment length=4}]
	
	\draw[fill=gray!50] (4-\x, 2-\x) rectangle (4+\x, 2+\x);
	\draw[fill=gray!50, rotate around={45:(\diag1,\diag1)}] (\diag1-\x, \diag1-\x) rectangle (\diag1+\x, \diag1+\x);
	\draw[fill=gray!50] (2-\x, 4-\x) rectangle (2+\x, 4+\x);
	\draw[fill=gray!50, rotate around={45:(\diag2,\diag1)}] (\diag2-\x, \diag1-\x) rectangle (\diag2+\x, \diag1+\x);
	
	\draw[fill=gray!50] (0-\x, 2-\x) rectangle (0+\x, 2+\x);
	\draw[fill=gray!50, rotate around={45:(\diag2,\diag2)}] (\diag2-\x, \diag2-\x) rectangle (\diag2+\x, \diag2+\x);
	\draw[fill=gray!50] (2-\x, 0-\x) rectangle (2+\x, 0+\x);
	\draw[fill=gray!50, rotate around={45:(\diag1,\diag2)}] (\diag1-\x, \diag2-\x) rectangle (\diag1+\x, \diag2+\x);
	
	\draw[spring] (0+\x, 2) -- (4-\x, 2);
	\draw[spring] (\diag2+0.7071*\x, \diag2+0.7071*\x) -- (\diag1-0.7071*\x, \diag1-0.7071*\x);
	\draw[spring] (2, 0+\x) -- (2, 4-\x);
	\draw[spring] (\diag1-0.7071*\x, \diag2+0.7071*\x) -- (\diag2+0.7071*\x, \diag1-0.7071*\x);
	
	\node at (4, 1.6) {$M$};
	
\end{circuitikz}

\caption{A diagram of the mass-spring system which is analogous to a non-rotating NS. In this particular example, we have 4 equally-spaced oscillators, each made up of a pair of masses, $M$, connected by a light spring. The oscillators are independent of one another and the masses are restricted to oscillate radially.}
\label{fig: oscillators}

\end{figure}
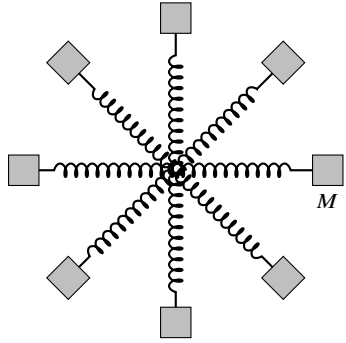

The position vector of a given mass can be written in terms of plane polar coordinates ($r, \phi$) as
\begin{align}
\bm{r}(\phi, t) = r(\phi, t) \bm{e}_r(\phi)~.
\end{align}
For a given oscillator, if one mass has position $\bm{r}$, the other has position $-\bm{r}$, so we need only keep track of one mass in each pair. We will consider a non-rotating array of masses, so that the angular location of a given mass does not change in time, i.e.~the masses just move in and out radially. In this particular set-up, the mass-spring system is analogous to oscillations on a non-rotating NS. One could straightforwardly add rotation if wanted.

We can set up the initial data to mimic a mode, such that
\begin{align}
\label{eq:initial_data}
r(\phi, t) = R + \alpha R \cos(m \phi + \omega t)~,
\end{align}
where $\alpha$ is a dimensionless number that gives the amplitude of the ``mode''. This oscillation has an associated pattern speed of
\begin{align}
\omega_\text{p} = - \frac{\omega}{m}~.
\end{align}
Only in the limit of there being infinitely many pairs of oscillators, such that there is a mass at every single value of $\phi$, will a perturbation of the above form really produce a rigidly rotating pattern. In what follows, we will consider a relatively small number of oscillators, so this notion of a rotating pattern is only approximate.

The radial velocity of one of the masses is
\begin{align}
\dot{r}(\phi, t) = - \alpha \omega R \sin(m \phi + \omega t)~,
\end{align}
so the total energy of each oscillator is 
\begin{align}
E_\text{osc} = 2 \times \frac{1}{2} M (\alpha \omega R)^2 = \alpha^2 \omega^2 M R^2~,
\end{align}
where the factor of 2 comes from the fact that each oscillator consists of a pair of masses. The total energy per single mass is the sum of the kinetic energy and elastic potential energy, but for a simple harmonic oscillator, these are always out of phase so that the total energy is always constant. This is why we can just use the maximum amplitude of the kinetic energy in the expression for the total energy per single mass.

The energy of $N$ oscillators is simply $N$ times this
\begin{align}
E = N E_\text{osc} = N \alpha^2 \omega^2 M R^2~.
\end{align}

We now move onto finding the angular momentum. However, we quickly find that there would not be any angular momentum for a system of $N$ oscillators. Each oscillator is a pair of masses oscillating in a straight line so carries no angular momentum, and if we have $N$ lots of them, irrespective of how they are arranged, there is still no angular momentum. It is here where the mass-spring analogue differs from a NS, as a propagating mode on a NS carries angular momentum. The difference arises because, for an oscillating NS, the motion of the fluid elements is not purely radial, in general. Nevertheless, this distinction does not make a difference in calculating the GW back-reaction.

We will now write down the equations that govern the rate of change of energy and angular momentum as a result of GW emission. In the weak field and slow motion approximation, the equations are given by the quadrupole formulae
\begin{align}
\dot{E} = \dot{E}_\text{GW} &= \frac{1}{5} \frac{G}{c^5} \left\langle \mathsout{\dddot{I}}_{ij} \mathsout{\dddot{I}}^{~ij} \right\rangle~, \\
\dot{J}^i = \dot{J}^i_\text{GW} &= \frac{2}{5} \frac{G}{c^5} \varepsilon^{ijk} \left\langle \mathsout{\ddot{I}}_{~j}^{\hspace{1pt}l} \mathsout{\dddot{I}}_{kl} \right\rangle~,
\end{align}
where $\mathsout{I}_{ij}$ is the trace-reduced mass quadrupole moment tensor and the dots represent time derivatives. To get $\mathsout{I}_{ij}$, we first calculate the mass quadrupole moment tensor, which is
\begin{align}
I_{ij}^\text{single} = M x_i x_j~,
\end{align}
for a single mass, where $x_i$ is the (time-dependent) position vector in Cartesian coordinates. Since the other mass in the pair has coordinates $-x_i$, the corresponding expression for an oscillator is twice as large, so
\begin{align}
\label{app_mass_quadrupole_osc}
I_{ij}^\text{osc} = 2 M x_i x_j~.
\end{align}
In trace-reduced form, we have
\begin{align}
\mathsout{I}_{ij}^\text{osc} = 2M (x_i x_j - \frac{1}{3} \delta_{ij} x^k x_k)~.
\end{align}
The total trace-reduced quadrupole moment tensor of the system is then the sum of all oscillators. Using $n$ to label the oscillators, we have
\begin{align}
\mathsout{I}_{ij} = \sum_{n=1}^{N} \mathsout{I}_{ij}^{(n), \text{osc}}~.	
\end{align}

We will now find the contribution to the quadrupole moment tensor from just one oscillator placed at an arbitrary angular position $\phi$. We will say $\phi=0$ corresponds to the 1-axis. By simple trigonometry, we write
\begin{align}
\label{app_x_1}
x_1 &= R\left[ 1 + \alpha \cos (m \phi + \omega t) \right] \cos\phi~, \\
x_2 &= R\left[ 1 + \alpha \cos (m \phi + \omega t) \right] \sin\phi~,
\end{align}
with $x_3 = 0$. Since we will only consider time derivatives of $\mathsout{I}_{ij}$, it would be useful to have the time derivative of $x_i$
\begin{align}
\dot{x}_1 &= - \alpha \omega R \sin (m \phi + \omega t) \cos\phi~, \\
\label{app_xdot_2}
\dot{x}_2 &= - \alpha \omega R \sin (m \phi + \omega t) \sin\phi~,
\end{align}
with $\dot{x}_3 = 0$. The time derivative of equation~(\ref{app_mass_quadrupole_osc}) is $\dot{I}_{ij}^\text{osc} = 2 M ( \dot{x}_i x_j + x_i \dot{x}_j )$ and after substituting in equations~(\ref{app_x_1})~--~(\ref{app_xdot_2}) and reducing by the trace, we get
\begin{align} 
\mathsout{\dot{I}}_{ij}^\text{osc} = - \frac{2}{3} \alpha \omega M R^2 \left\{ \sin(m \phi + \omega t) + \frac{1}{2} \alpha \sin[2(m \phi + \omega t)] \right\} \left(~
\begin{matrix}
1 + 3\cos(2\phi) & 3\sin(2\phi) & 0 \\ 
3\sin(2\phi) & 1 - 3\cos(2\phi) & 0\\
0 & 0 & -2 \\
\end{matrix}
~\right)~.
\end{align}
for an oscillator at angle $\phi$. From this, one can see that there would be two frequencies of GWs given off from an oscillating pair of masses, at $\omega$ and $2\omega$. The $2\omega$ radiation is weaker by a factor of $\alpha$ and, in terms of the GW luminosity, would represent an $\alpha^4$ term. On the other hand, the $\omega$ radiation leads to the leading order GW luminosity term which is on the order of $\alpha^2$. It is safe to no longer consider the $2\omega$ radiation but we will explicitly include a big-$\mathcal{O}$ term whenever a term derived from the $2\omega$ radiation comes up.

Now that we have $\mathsout{\dot{I}}_{ij}$ for one oscillator, we can add multiple together to create a ring of masses. In the case where $N = 2$, we would get orthogonal oscillators that oscillate out of phase with each other (recall the initial data of equation~(\ref{eq:initial_data})). It is clear that we would not be able to get a sense of the direction the mode would be propagating, making this particular example not suitable for our query about the angular momentum. Although $N=3$ should suffice, we will go to the $N=4$ case as the trigonometric functions simplify greatly. 

We therefore place oscillators at $\phi = 0, \frac{\pi}{4}, \frac{\pi}{2}, \frac{3\pi}{4}$ such that
\begin{align}
\mathsout{\dot{I}}_{ij} = \mathsout{\dot{I}}_{ij}^\text{osc}(\phi = 0) + \mathsout{\dot{I}}_{ij}^\text{osc}(\phi = \frac{\pi}{4}) + \mathsout{\dot{I}}_{ij}^\text{osc}(\phi = \frac{\pi}{2}) + \mathsout{\dot{I}}_{ij}^\text{osc}(\phi = \frac{3\pi}{4})~,
\end{align}
for the entire 4-oscillator system. Differentiating an appropriate number of times and putting it into the quadrupole formulae for $\dot{E}_\text{GW}$ and $\dot{J}_\text{GW}$, after some algebra, we find
\begin{align}
\dot{E} &= \frac{1}{10} \frac{G}{c^5} \alpha^2 \omega^6 M_\text{tot}^2 R^4 + \mathcal{O}(\alpha^4)~, \\
\label{app_GW_angular_momentum}
\dot{J} &= \mp \frac{1}{5} \frac{G}{c^5} \alpha^2 \omega^5 M_\text{tot}^2 R^4~,
\end{align}
for $m=\pm2$ and where $M_\text{tot} = 8M$. We set $\dot{J}^3 = \dot{J}$ since angular momentum is only lost along the 3-axis for this system. Also, it should be noted that even with the $2\omega$ radiation, $\dot{J}$ only has a term that comes from the $\omega$ radiation.

Interestingly, if we work backwards and replace $\pm 2$ with $m$, we find that $\dot{E}$ and $\dot{J}$ are related by the pattern speed
\begin{align}
\dot{E} = - \frac{\omega}{m} \dot{J} = \omega_\text{p} \dot{J}~,
\end{align}
which is reminiscent of an equation in \citet{friedmanschutz1978b} that comes immediately after their equation~(71). See also equation~(\ref{eq:E-dot_delta_J-dot_relation}) in the main text of this paper. 

There are now two facts that we can use to argue that a GW back-reaction must exist. Initially, there is no angular momentum in the system, since each oscillator is forced to oscillate radially. Moreover, we have just shown that GWs carrying angular momentum will be emitted as a result of a mode propagating around our mass-spring system. Given the masses oscillate radially, one is forced to conclude that the array of masses starts collectively rotating in the opposite direction of the mode in order to conserve angular momentum. This is the GW back-reaction that we set out to expose.

As for how much this affects the array of masses, let us write down the total moment of inertia of the system
\begin{align}
I = 8 M R^2 = M_\text{tot} R^2~,
\end{align}
and since the emission of angular momentum through GWs is $\dot{J}$, there must be an equal and opposite torque back on the system, $-\dot{J}$. This means there will be an angular acceleration of
\begin{align}
\dot{\Omega} = - \frac{\dot{J}}{I} = \frac{1}{10} \frac{G}{c^5} m \alpha^2 \omega^5 M_\text{tot} R^2~,
\end{align}
c.f. equation~(\ref{back_reaction_2}). One can see that for a prograde mode, $m < 0$, so $\dot{\Omega}>0$ and for a retrograde mode, $m > 0$, so $\dot{\Omega} < 0$. In words, the propagation of a prograde mode causes the array of masses to collectively rotate in the retrograde direction, and the opposite is true for a retrograde mode.

\section{Real and complex mode eigenfunctions}
\label{appendix: real and complex mode eigenfunctions}

In their papers on Lagrangian perturbation theory for rotating stars, \citet{friedmanschutz1978, friedmanschutz1978b} wrote down some of their results for the canonical energy, $E_\text{c}$, and canonical angular momentum, $J_\text{c}$, in terms of a purely real Lagrangian displacement vector $\xi$. At other times, they used complex $\xi$ to calculate the same quantities. This appendix shows how to switch back-and-forth between the two descriptions -- it is not as simple as treating real $\xi$ as being the real part of complex $\xi$. We will refer to equations from \citet{friedmanschutz1978} with (FSa...).

For clarity, we will use $\xi$ to denote real eigenfunctions, with $E_\text{c}$ being the canonical energy calculated using equations from \citet{friedmanschutz1978} specific to $\xi$. Similarly, we will use $\tilde{\xi}$ to denote complex eigenfunctions, with $\tilde{E}_\text{c}$ being the canonical energy calculated using equations from \citet{friedmanschutz1978} specific to $\tilde{\xi}$. Here, we only focus on the canonical energy, which is sufficient for our proof, but one can do the same calculation for the canonical angular momentum and yield the same result.

The following calculation is based on the idea that regardless of whether we calculate the canonical energy from real or complex methods, we must have the same final value, i.e.~$E_\text{c} = \tilde{E}_\text{c}$. The stars we look at here are general, i.e.~compressible and rotating, but one can take appropriate limits for the incompressible and/or non-rotating case, which we have checked remains valid.

We begin with finding $E_\text{c}$. The Lagrangian density, $\mathcal{L}$, is given by
\begin{align}
\label{app_langrangian_density}
\mathcal{L} = \frac{1}{2} \left( \dot{\xi}_i A^i_{~j} \dot{\xi}^j + \dot{\xi}_i B^i_{~j} \xi^j - \xi_i C^i_{~j} \xi^j \right)~,
\end{align}
(FSa35), which we partially differentiate with respect to $\dot{\xi}^k$ to get the momentum conjugate
\begin{align}
\label{app_momentum_conjugate}
\frac{\partial \mathcal{L}}{\partial \dot{\xi}^k} = g_{ik} \left( A^i_{~j} \dot{\xi}^j + \frac{1}{2} B^i_{~j} \xi^j \right)~,
\end{align}
where $g_{ik}$ is the metric tensor. Then, we apply the Euler-Lagrange equation 
\begin{align}
\frac{d}{dt} \left( \frac{\partial \mathcal{L}}{\partial \dot{\xi}^k} \right) = \frac{\partial \mathcal{L}}{\partial \xi^k}~,
\end{align}
to find the equation of motion
\begin{align}
\label{app_equation_of_motion}
A^i_{~j} \ddot{\xi}^j + B^i_{~j} \dot{\xi}^j + C^i_{~j} \xi^j = 0~,
\end{align}
(FSa15) where $A$ and $C$ are symmetric operators and $B$ is an anti-symmetric operator. $E_\text{c}$ can be calculated with its definition
\begin{align}
E_\text{c} \equiv \int_{V} \left( \dot{\xi}^k \frac{\partial \mathcal{L}}{\partial \dot{\xi}^k} - \mathcal{L} \right) dV~,
\end{align}
(FSa44) and after inputting equations~(\ref{app_langrangian_density}) and (\ref{app_momentum_conjugate}), one finds
\begin{align}
E_\text{c} = \frac{1}{2} \int_{V} \left( \dot{\xi}_i A^i_{~j} \dot{\xi}^j + \xi_i C^i_{~j} \xi^j \right) dV~.
\end{align}
Note that, when written in this form, $E_\text{c}$ is the same as its $B=0$ (non-rotating) equivalent. However, the dependence on $B$ is exposed when we use equation~(\ref{app_equation_of_motion}) to eliminate $C$
\begin{align}
\label{app_canonical_energy_integral_B}
E_\text{c} = \frac{1}{2} \int_{V} \left( \dot{\xi}_i A^i_{~j} \dot{\xi}^j - \xi_i A^i_{~j} \ddot{\xi}^j - \xi_i B^i_{~j} \dot{\xi}^j \right) dV~.
\end{align}
If real $\xi$ is defined by
\begin{align}
\xi^i(r, \theta, \phi, t) \equiv \zeta^i(r,\theta) \cos(m \phi + \omega t + \Phi_{(i)})~,
\end{align}
where $\zeta^i$ represents the amplitude and $\Phi_{(i)}$ is a constant (whose index is not to be summed over), then the time derivatives are
\begin{align}
\dot{\xi}^i &= - \omega \zeta^i(r,\theta) \sin(m \phi + \omega t + \Phi_{(i)})~, \\
\ddot{\xi}^i &= - \omega^2 \zeta^i(r,\theta) \cos(m \phi + \omega t + \Phi_{(i)})~.
\end{align}
Substituting into equation~(\ref{app_canonical_energy_integral_B}), we find
\begin{align}
E_\text{c} = \frac{1}{2} \int_{V} \left[ \omega^2 \zeta_i A^i_{~j} \zeta^j \cos(\Phi_{(i)} - \Phi_{(j)}) + \omega \zeta_i B^i_{~j} \zeta^j \cos(m \phi + \omega t + \Phi_{(i)}) \sin(m \phi + \omega t + \Phi_{(j)}) \right] dV~,
\end{align}
where we have made use of the trigonometric identity, $\cos(A-B) \equiv \cos A \cos B + \sin A \sin B$. We now take a look at the second term and sum over a pair of elements whilst taking advantage of the anti-symmetric properties of $B$ and the sine function, e.g.
\begin{align}
(1,2) + (2,1) &= \zeta_1 B^1_{~2} \zeta^2 \cos(m \phi + \omega t + \Phi_{(1)}) \sin(m \phi + \omega t + \Phi_{(2)}) + \zeta_2 B^2_{~1} \zeta^1 \cos(m \phi + \omega t + \Phi_{(2)}) \sin(m \phi + \omega t + \Phi_{(1)})~, \\
&= \zeta_1 B^1_{~2} \zeta^2 \left[ \cos(m \phi + \omega t + \Phi_{(1)}) \sin(m \phi + \omega t + \Phi_{(2)}) - \cos(m \phi + \omega t + \Phi_{(2)}) \sin(m \phi + \omega t + \Phi_{(1)}) \right]~, \\
&= \zeta_1 B^1_{~2} \zeta^2 \sin(\Phi_{(2)} - \Phi_{(1)})~, \\
&= - \frac{1}{2} \zeta_1 B^1_{~2} \zeta^2 \sin(\Phi_{(1)} - \Phi_{(2)}) - \frac{1}{2} \zeta_1 B^1_{~2} \zeta^2 \sin(\Phi_{(1)} - \Phi_{(2)})~, \\
&= - \frac{1}{2} \zeta_1 B^1_{~2} \zeta^2 \sin(\Phi_{(1)} - \Phi_{(2)}) - \frac{1}{2} \zeta_2 B^2_{~1} \zeta^1 \sin(\Phi_{(2)} - \Phi_{(1)})~,
\end{align}
which can be generalised to all pairs meaning the full summation is
\begin{align}
\zeta_i B^i_{~j} \zeta^j \cos(m \phi + \omega t + \Phi_{(i)}) \sin(m \phi + \omega t + \Phi_{(j)}) = -\frac{1}{2} \zeta_i B^i_{~j} \zeta^j \sin(\Phi_{(i)} - \Phi_{(j)})~.
\end{align}
Therefore, our final expression for $E_\text{c}$ is
\begin{align}
\label{app_E_c_real}
E_\text{c} = \frac{1}{2} \int_{V} \left[ \omega^2 \zeta_i A^i_{~j} \zeta^j \cos(\Phi_{(i)} - \Phi_{(j)}) -\frac{1}{2} \omega \zeta_i B^i_{~j} \zeta^j \sin(\Phi_{(i)} - \Phi_{(j)}) \right] dV~.
\end{align}
Now, we move onto calculating $\tilde{E}_\text{c}$ with $\tilde{\xi}$. First, we define the complex inner product
\begin{align}
\langle \tilde{\xi}, \tilde{\eta} \rangle \equiv \int_{V} (\tilde{\xi}_i)^* \tilde{\eta}^i dV~,
\end{align}
(above FSa36) which allows us to write $\tilde{E}_\text{c}$ as
\begin{align}
\tilde{E}_\text{c} = \frac{1}{2} \langle \dot{\tilde{\xi}}, A \dot{\tilde{\xi}} \rangle + \frac{1}{2} \langle \tilde{\xi}, C \tilde{\xi} \rangle = \frac{1}{2} \int_{V} \left( \dot{\tilde{\xi}}^*_i A^i_{~j} \dot{\tilde{\xi}}^j + \tilde{\xi}^*_i C^i_{~j} \tilde{\xi}^j \right) dV ~,
\end{align} 
(FSa43) where now operators $A$ and $C$ are Hermitian and operator $B$ is anti-Hermitian. Again, eliminating $C$ using the equation of motion, we get
\begin{align}
\label{app_canonical_energy_complex_B}
\tilde{E}_\text{c} = \frac{1}{2} \int_{V} \left( \dot{\tilde{\xi}}^*_i A^i_{~j} \dot{\tilde{\xi}}^j - \tilde{\xi}^*_i A^i_{~j} \ddot{\tilde{\xi}}^j - \tilde{\xi}^*_i B^i_{~j} \dot{\tilde{\xi}}^j \right) dV~.
\end{align} 
For complex modes, we define
\begin{align}
\tilde{\xi}^i(r, \theta, \phi, t) \equiv \tilde{\zeta}^i(r,\theta) e^{i(m \phi + \omega t)} \equiv |\tilde{\zeta}^i(r,\theta)| e^{i(m \phi + \omega t + \Phi_{(i)})} ~,
\end{align}
where we used $\tilde{\zeta}^i(r,\theta) = |\tilde{\zeta}^i(r,\theta)| e^{i\Phi_{(i)}}$. Differentiating with respect to time, we get
\begin{align}
\dot{\tilde{\xi}}^i &= i \omega \tilde{\xi}^i ~~~ \rightarrow ~~~ (\dot{\tilde{\xi}}^{i})^* = - i \omega (\tilde{\xi}^{i})^* ~, \\
\ddot{\tilde{\xi}}^i &= - \omega^2 \tilde{\xi}^i ~,
\end{align}
which can be substituted into equation~(\ref{app_canonical_energy_complex_B}) to give
\begin{align}
\label{app_canonical_energy_complex_exponentials}
\tilde{E}_\text{c} = \int_{V} \left( \omega^2 \tilde{\xi}^*_i A^i_{~j} \tilde{\xi}^j - \frac{1}{2} i \omega \tilde{\xi}^*_i B^i_{~j} \tilde{\xi}^j \right) dV = \int_{V} \left( \omega^2 |\tilde{\zeta}_i| A^i_{~j} |\tilde{\zeta}^j| e^{-i(\Phi_{(i)} - \Phi_{(j)})} - \frac{1}{2} i \omega |\tilde{\zeta}_i| B^i_{~j} |\tilde{\zeta}^j| e^{-i(\Phi_{(i)} - \Phi_{(j)})} \right) dV ~.
\end{align} 
Like before, we consider a pair of elements belonging to the summation over $i$ and $j$. Using the Hermitian properties of $A$, we evaluate the first term with
\begin{align}
(1,2) + (2,1) &= |\tilde{\zeta}_1| A^1_{~2} |\tilde{\zeta}^2| e^{-i(\Phi_{(1)} - \Phi_{(2)})} + |\tilde{\zeta}_2| A^2_{~1} |\tilde{\zeta}^1| e^{-i(\Phi_{(2)} - \Phi_{(1)})}~, \\
&= |\tilde{\zeta}_1| |\tilde{\zeta}^2| \left\{ A^1_{~2} e^{-i(\Phi_{(1)} - \Phi_{(2)})} + \left[A^1_{~2} e^{-i(\Phi_{(1)} - \Phi_{(2)})} \right]^*~\right\}~, \\
&= 2 |\tilde{\zeta}_1| A^1_{~2} |\tilde{\zeta}^2| \cos(\Phi_{(1)} - \Phi_{(2)}) ~, \\
&= |\tilde{\zeta}_1| A^1_{~2} |\tilde{\zeta}^2| \cos(\Phi_{(1)} - \Phi_{(2)}) + |\tilde{\zeta}_1| A^1_{~2} |\tilde{\zeta}^2| \cos(\Phi_{(1)} - \Phi_{(2)}) ~, \\
&= |\tilde{\zeta}_1| A^1_{~2} |\tilde{\zeta}^2| \cos(\Phi_{(1)} - \Phi_{(2)}) + |\tilde{\zeta}_2| A^2_{~1} |\tilde{\zeta}^1| \cos(\Phi_{(2)} - \Phi_{(1)}) ~,
\end{align}
so the full summation is
\begin{align}
|\tilde{\zeta}_i| A^i_{~j} |\tilde{\zeta}^j| e^{-i(\Phi_{(i)} - \Phi_{(j)})} = |\tilde{\zeta}_i| A^i_{~j} |\tilde{\zeta}^j| \cos(\Phi_{(i)} - \Phi_{(j)})~.
\end{align} 
One can do the same calculation but for the second term in equation~(\ref{app_canonical_energy_complex_exponentials}) and one would find
\begin{align}
|\tilde{\zeta}_i| B^i_{~j} |\tilde{\zeta}^j| e^{-i(\Phi_{(i)} - \Phi_{(j)})} = - i |\tilde{\zeta}_i| B^i_{~j} |\tilde{\zeta}^j| \sin(\Phi_{(i)} - \Phi_{(j)})~.
\end{align} 
Putting these into equation~(\ref{app_canonical_energy_complex_exponentials}), we get the final result of
\begin{align}
\tilde{E}_\text{c} = \int_{V} \left( \omega^2 |\tilde{\zeta}_i| A^i_{~j} |\tilde{\zeta}^j| \cos(\Phi_{(i)} - \Phi_{(j)}) - \frac{1}{2} \omega |\tilde{\zeta}_i| B^i_{~j} |\tilde{\zeta}^j| \sin(\Phi_{(i)} - \Phi_{(j)}) \right) dV ~.
\end{align} 
Comparing this with $E_\text{c}$ from equation~(\ref{app_E_c_real}), we see that we must have
\begin{align}
|\tilde{\zeta}_i| |\tilde{\zeta}^j| = \frac{1}{2} \zeta_i \zeta^j ~~~ \rightarrow ~~~ |\tilde{\zeta}^i| = \frac{1}{\sqrt{2}} \zeta^i~,
\end{align}
to have the same canonical energy from real and complex methods. In other words, one must normalise the complex eigenfunctions, $\tilde{\xi}$, by a factor of $\frac{1}{\sqrt{2}}$.

\bsp	
\label{lastpage}
\end{document}